\documentclass[pdftex,twocolumn,epjc3]{svjour3}          

\RequirePackage[T1]{fontenc}

\smartqed  

\RequirePackage{graphicx}
\RequirePackage{mathptmx}      
\RequirePackage{flushend}
\RequirePackage[numbers,sort&compress]{natbib}
\RequirePackage[colorlinks,citecolor=blue,urlcolor=blue,linkcolor=blue]{hyperref}

\usepackage[dvipsnames]{xcolor}

\newcommand{\be}{\begin{equation}}
\newcommand{\ee}{\end{equation}}
\newcommand{\bea}{\begin{eqnarray}}
\newcommand{\eea}{\end{eqnarray}}

\usepackage{multirow}
\usepackage{soul}
\usepackage{float}
\usepackage{graphicx}
\usepackage{times}
\DeclareUnicodeCharacter{2212}{-}

\begin{document}

\title{High-Density behavior of symmetry energy and speed of sound in the dense matter within an effective chiral model}


\author{Prashant Thakur\thanksref{e1,addr1},
        N. K. Patra\thanksref{e2,addr1},
        T. K. Jha\thanksref{e3,addr1} 
        \and
        Tuhin Malik \thanksref{e4,addr2}     
}

\thankstext{e1}{e-mail: p20190072@goa.bits-pilani.ac.in}
\thankstext{e2}{e-mail: nareshkumarpatra3@gmail.com}
\thankstext{e3}{e-mail: tkjha@goa.bits-pilani.ac.in}
\thankstext{e4}{e-mail: tm@uc.pt}

\institute{Department of Physics, BITS-Pilani, K. K. Birla Goa
Campus, Goa 403726, India\label{addr1}
          \and
          CFisUC, Department of Physics, University of Coimbra,
3004-516 Coimbra, Portugal\label{addr2}
}

\date{Received: date / Accepted: date}

\maketitle

\begin{abstract}
With an effective chiral model, we investigate how the mesonic cross couplings $\sigma-\rho$ and $\omega-\rho$ affect the density content of the symmetry energy and its higher-order slope parameters. Earlier mentioned cross-couplings are crucial to controlling the density content of symmetry energy. For this purpose, we did a case study for different values of the symmetry energy $J_1$, defined at density 0.1 fm$^{-3}$ in the range (23.4 - 25.2) for a fixed value of the slope of the symmetry energy $L_0 = 60$ MeV at saturation density and investigate its effect on the higher-order coefficients and their influence on the underlying equation of state. We found that the model with $J_1= 24.6$ MeV is more favorable with the pure neutron matter (PNM) constraints obtained from $\chi$EFT calculations. In addition, we show that all of our models predict a monotonically increasing speed of sound up to four times the saturation density. The speed of sound decreases/saturates above that point and approaches the conformal limit approximately $\sqrt{1/3}~c$ at the center of the maximum mass star.
\end{abstract}

\section{Introduction}
Constraining the equation of state (EOS) of dense nuclear matter, supported by astrophysical observations \cite{Lattimer:2000nx, Watts:2016uzu, Ozel:2016oaf, Oertel:2016bki, Vidana:2018bdi, Bombaci:2018ksa}, and terrestrial experiments \cite{Danielewicz:2002pu, Baran:2004ih, Li:2008gp, Tsang:2019mlz, Trautmann:2012nk, Lopez:2017ctx, Giuliani:2013kna, Garg:2018uam, Ono:2019jxm} remains one of the primary focus in dense matter studies for decades. A dense matter EOS can consist of several phases or compositions, including hyperons, quarks, superconducting matter, and colored superconducting matter. The density-dependence of symmetry energy plays an important role in regulating the particle content in the underlying EOS, particularly in studies of highly asymmetric matter such as those speculated in the core of neutron stars. It is to be noted that the symmetry energy slope parameters can be equivalently expressed as a series expansion in matter density. Here, in addition to $J_0$, the symmetry energy defined at nuclear saturation density ($\rho_0 \sim  0.153 ~$fm$^{-3}$), its slope parameter $L_0$, the curvature parameter $K_{\rm sym,0}$, and the skewness parameter $Q_{\rm sym,0}$, as well as other higher-order terms may dictate the behavior of a particular EOS at higher densities. Based on the current knowledge of nuclear masses and giant dipole polarizability, the symmetry energy has been constrained $J_0 \simeq 32.25\pm 2.5$ MeV \cite{Roca-Maza:2013mla, Roca-Maza:2015eza, Vinas:2013hua, Mondal:2016roo} and it's slope parameter $L_0 \simeq 58.9 \pm 16 $ MeV \cite{Li:2014qta, Trippa:2008gr, Moller:2012pxr, Tsang:2012se, Lattimer:2012xj, Oertel:2016bki} at nuclear saturation density.
However, there are very few known theoretical constraints on $K_{\rm sym,0}$ and constraints on $Q_{\rm sym,0}$ \cite{Lattimer:2012xj, Lattimer:2023rpe, Malik:2022zol} and therefore, one of our objectives here is to analyze their  interdependence with a model based on microphysics. Numerous relativistic mean-field models (RMF) \cite{Walecka:1974qa, Serot1986, Gambhir:1990uyn, Patra:1991wy, Serot:2002ei} are known to have been successfully applied theoretically to extract finite nuclear properties across the periodic table as well as in applications of nuclear matter. Broadly, in these theories, nuclear interactions have been expressed in terms of the scalar meson $\sigma$, the vector mesons $\omega$ and the iso-vector meson $\rho$ and their cross-couplings and the respective fields being expressed in mean-field where quantum fluctuations are ignored. The formalism seems to be more so valid when the density or the source terms are large \cite{Serot1986, Todd-Rutel:2005yzo, Agrawal:2010wg, Kumar:2017xdy}.

Analogously, models based on chiral symmetry, introduced by Gell-Mann $\&$ Levy \cite{Gell-Mann:1960mvl} were subsequently applied to nuclear matter studies \cite{Lee:1974ma}; however, had limited applications in the finite nuclear domain \cite{Ogawa:2003ns, Chanfray:2006nz}. One of the major drawbacks of the model is that at short distances or high-momentum transfer, the interactions become weaker \cite{Thomas:1982kv}. The vacuum jumps to a chirally restored abnormal vacuum (Lee-Wick vacuum) \cite{Lee:1974ma, Lee:1974uu}. The problem \cite{Thomas:2004iw} may be overcome by including logarithmic terms of the scalar field in chiral potentials \cite{Furnstahl:1993wx, Serot:2002ei, Heide:1993yz, Mishustin:1993ub, Papazoglou:1998vr} which intercept the normal vacuum from collapsing which may help in describing the finite nuclear properties \cite{Schramm:2002xi, Tsubakihara:2006se, Tsubakihara:2009zb}. Subsequent inclusion of the dynamically generated mass of the vector meson in the model \cite{Boguta:1983uz, Sahu:1993db} reported an unrealistic high nuclear incompressibility ($K$) value, which was taken care of later by introducing higher-order terms of scalar meson field and applied to nuclear matter studies at both low \cite{Sahu2004} and high densities and neutron stars  \cite{Jha:2008yth, Jha2006, Jha2010}. Lately, in order to improve the density content of the symmetry energy, mesonic cross-couplings were incorporated \cite{Malik2017} and applied to study magnetized neutron stars \cite{Patra2020}. Moreover, the higher-order interactions in the chiral fields are desirable as it is known to mimic the three-body forces, which may play a vital role in dense matter studies.

In the present work, we focus on the aspects of the isospin part of nuclear interaction and analyze its dependence on the in-medium nuclear interactions and the resulting equation of state (EOS), particularly on the density dependence of the nuclear symmetry energy \cite{Danielewicz:2002pu}. 
Although the nuclear symmetry energy at normal matter density is fairly well known, however, they are poorly known or known with large uncertainties at supra-normal densities \cite{Li:1997px}.  Our primary objective is to study the behavior of different order slope parameters of the nuclear symmetry energy up to larger densities. For this, we employ the effective chiral model with mesonic cross interaction such as $\sigma-\rho$ and $\omega-\rho$. There are three parameters in our model regulating the density content of symmetry energy. We systematically study the correlations and the effect of those parameters on different slope parameters of the nuclear symmetry energy.

 It was often shown by several authors that the speed of sound has exhibited a maximum at an energy density of approximately 500 MeV fm$^{-3}$ or three times the saturation density when the EOS developed by physics-agnostic perturbative QCD calculations is conditioned at high densities \cite{Kurkela:2022elj, Gorda:2022jvk}. A similar phenomenon, however, does not occur when only astronomical restrictions are applied. Accordingly, we also investigate the behavior of sound up to higher densities.

The paper is organized as follows. In Section \ref{sec2}, we present the details of the hadronic model. The results of our calculations are discussed in Section \ref{sec3}. Section \ref{sec4} contains the summary and conclusions of the present work.

\section{The equation of state}\label{sec2} 
The effective chiral lagrangian is as given in Eq. (\ref{Lag_density1}), where the hadronic degrees of freedom are the pseudo-scalar meson $\pi$, the scalar meson $\sigma$, the vector meson $\omega$ and the isovector $\rho-$meson \cite{Malik2017}.
\small{ 
\begin{eqnarray}
\label{Lag_density1}
  {\cal{L}}&&= \bar{\psi}_{B}\Big[\left(i \gamma_{\mu}\partial^{\mu}
        -g_{\omega}\gamma_{\mu}\omega^{\mu}-\frac{1}{2}g_{\rho}\vec{\rho_{\mu}}. \vec{\tau}\gamma^{\mu}\right)
     -g_{\sigma} (\sigma + i\gamma_{5}\vec{\tau}.\vec{\pi})\Big]\psi_{B} \nonumber\\
     &&+ \frac{1}{2}\left( \partial_{\mu}\vec{\pi}.\partial^{\mu}\vec{\pi}
    + \partial_{\mu}\sigma \partial^{\mu}\sigma \right)        
     - \frac{\lambda}{4}\left(x^{2}-x_{0}^{2}\right)^{2}- \frac{\lambda b} {6 m^{2}}(x^{2}-x_{0}^{2})^{3}  \nonumber \\ 
    &&-\frac{\lambda c}{8 m^{4}}(x^{2}-x_{0}^{2})^{4}
     - \frac{1}{4} F_{\mu\nu}F^{\mu\nu}+ \frac{1}{2}g_{ \omega}^{2}x^{2}\left(\omega_{\mu} \omega^{\mu}\right) 
    - \frac{1}{4} \vec{R_{\mu \nu}}. \vec{R^{\mu \nu}} \nonumber \\ 
    &&+ \frac{1}{2} 
    {m_\rho^\prime}^2\vec{\rho_{\mu}}.\vec{\rho^{\mu}} 
    +\eta_1 \left(\frac{1}{2} g_\rho^2x^2 \vec{\rho_{\mu}}.\vec{\rho^{\mu}} \right) 
    + \eta_2 \left(\frac{1}{2} g_{\rho}^2 \vec{\rho_{\mu}}.\vec{\rho^{\mu}} \omega_\mu \omega^\mu \right).
\end{eqnarray}}\normalsize
In the lagrangian above, $\psi_B$ is the nucleon iso-spin doublet interacting with the mesons. The $\frac{1}{4} F_{\mu\nu}F^{\mu\nu}$ and $\frac{1}{4} \vec{R_{\mu \nu}}\vec{R^{\mu \nu}}$ are kinetic terms for $\omega$ and $\rho$ respectively, where $F_{\mu\nu}=\partial_\mu \omega_\nu - \partial_\nu \omega_\mu$ and $\vec{R_{\mu \nu}} = \partial_\mu \rho_\nu - \partial_\nu \rho_\mu$.
The coupling strength for higher order scalar fields are $b$ and $c$,  respectively, and $\gamma^{\mu}$ and $\tau$ are the Dirac matrices and Pauli matrices, respectively. The interaction terms and higher order terms are given in terms of the chiral invariant field $x^2 = (\pi^2 + \sigma^2)$. The last two terms contain mesonic cross-coupling between $\rho$-$\sigma$ and $\rho$-$\omega$ whose coupling strengths are  $\eta_1$ and $\eta_2$, respectively. The vacuum expectation value of the scalar field is $x_0$ and the mass of the nucleon ($m$), the scalar ($m_{\sigma}$), and the vector meson mass ($m_{\omega}$), are related to $x_0$ through

\begin{eqnarray}
\label{ssb_mass}
m = g_{\sigma} x_0,~~ m_{\sigma} = \sqrt{2\lambda} x_0,~~
m_{\omega} = g_{\omega} x_0\ ,
\end{eqnarray}
where, $\lambda=\frac{(m_{\sigma}^2 -m_{\pi}^2)}{2 f_{\pi}^2}$, where and $f_\pi$ is the pion decay constant. In the mean-field treatment, we ignore the pion field and also set $m_\pi=0$. 
The derivation of the equation of motion of the meson fields and its EOS $(\varepsilon~\&~p)$ for the present model can be found in Ref. \cite{Malik2017}. In terms of $Y = x/x_0 = m^*/m$ and the corresponding couplings $C_\sigma\equiv g_\sigma^2/m_\sigma^2$ for scalar, $C_\omega\equiv g_\omega^2/m_\omega^2$ for vector and $C_\rho\equiv g_\rho^2/m_\rho^2$ for the iso-vector the energy density $(\epsilon)$ and pressure $(p)$ of the present model for a given baryon density can be calculated as,
 
 \begin{eqnarray}
\label{EOS_energy}
\epsilon && = \frac{1}{\pi^2}\sum_{i=n,p}\int_0^{k_F^i}k^2\sqrt{k^2+m^*{^2}}dk 
      + \frac{m^2}{8 C_{\sigma}}(1-Y^2)^2 \nonumber\\
      &&- \frac{b}{12 C_{\sigma}C_{\omega}}(1-Y^2)^3 
       + \frac{c}{16 m^2 C_{\sigma}C_{\omega}^2}(1-Y^2)^4 
      + \frac{1}{2}m_{\omega}^2\omega_{0}^2Y^2 \nonumber\\             
      &&+ \frac{1}{2}m_\rho^2\Big[1- \eta_1(1-Y^2) (C_\rho/C_\omega) 
      + 3 \eta_2 C_\rho \omega_0^2\Big] (\rho_3^0)^2, \\
\label{EOS_pressure}
   p &&= \frac{1}{3 \pi^2}\sum_{i=n,p}\int_0^{k_F^i}\frac{k^4}{\sqrt{k^2+m^*{^2}}}dk - \frac{m^2}{8 C_{\sigma}}(1-Y^2)^2  \nonumber \\
     && + \frac{b}{12 C_{\sigma}C_{\omega}}(1-Y^2)^3
     - \frac{c}{16 m^2 C_{\sigma}C_{\omega}^2}(1-Y^2)^4+ \frac{1}{2}m_{\omega}^2\omega_{0}^2Y^2 \nonumber\\
     && + \frac{1}{2}m_\rho^2 \Big[1- \eta_1(1-Y^2) (C_\rho/C_\omega) 
     + \eta_2 C_\rho \omega_0^2\Big] (\rho_3^0)^2.
\end{eqnarray}
In the equation above, $k_F^i$ is the fermi momentum of a nucleon and $\gamma$ is the spin degeneracy factor ($\gamma=4$ for Symmetric Nuclear Matter (SNM)). The parameters $C_\sigma$ and $C_\omega$ can be obtained by solving the field equations self-consistently tuned to satisfy standard nuclear matter saturation properties. The detailed calculation of the same can be found in our previous work, where it was evident that the couplings are very sensitive to the value of the effective mass \cite{Jha:2008yth}. Accordingly, the symmetry energy $S(\rho)$ in the present model using cross interactions (Eq. (\ref{Lag_density1})) is given as,

 \begin{eqnarray}
\label{sym_e}
   S(\rho) &=& \frac{k_F^2}{6 \sqrt{k_F^2 + m^*{^2}}}+ \frac{C_\rho k_F^3}
                   {12 \pi^2 (m^*_\rho/m_\rho)^2} + \frac{\eta_2 C_\rho^2 \omega_0^2 k_F^3 }{6 \pi^2 (m^*_\rho/m_\rho)^4 } \nonumber \\
           && - \frac{2 \eta_2 C_\rho^2 C_\omega k_F^9}{27 \pi^6 
        m_\omega^2 Y^4 (m^*_\rho/m_\rho)^4},
\end{eqnarray} 
where, ${m^*_\rho}^2=m_\rho^2\Big[1- \eta_1 (1-Y^2) (C_\rho/C_\omega)+ \eta_2 C_\rho \omega_0^2\Big]$ and $k_F= (3 \pi^2 \rho/ 2)^{1/3}$. The three coupling parameters $C_\rho$, $\eta_1$ and $\eta_2$ are fixed numerically for given values of symmetry energy at 0.1 fm$^{-3}$ $J_1$, symmetry energy $J_0$ and its slope parameter $L_0$ at saturation density. In general, the symmetry energy can be expressed in Taylor series expansion around saturation density $(\rho_0)$ as \cite{Dong:2015vga}

\bea
\label{sym_expan}
S(\rho)&&= J_0 + L_0 (\frac{\rho - \rho_0}{3 \rho_0}) + \frac{1}{2}K_{\rm sym,0}(\frac{\rho - \rho_0}{3 \rho_0})^2\nonumber\\   
&&+\frac{1}{6}Q_{\rm sym,0}(\frac{\rho - \rho_0}{3 \rho_0})^3
 + \frac{1}{24}Z_{\rm sym,0}(\frac{\rho - \rho_0}{3 \rho_0})^4\nonumber\\
 &&+ \mathcal O(\frac{\rho - \rho_0}{3 \rho_0})^5,
\eea
where $J_0$ is the symmetry energy coefficient, its slope parameter $L_0$, symmetry energy curvature parameter $K_{\rm sym,0}$ and the third order density derivatives of symmetry energy $Q_{\rm sym,0}$, all defined at $\rho_0$ \cite{Lattimer:2012xj}. The density dependence of all those parameters can be defined as,

\bea
L &=& 3 \rho_0 \frac{\partial S(\rho)}{\partial \rho} \label{eq-l0}, \\
K_{\rm sym} &=& 9 \rho_0^2 \frac{\partial^2 S(\rho)}{\partial\rho^2}\label{eq-ksym0}, \\
Q_{\rm sym} &=& 27 \rho_0^3 \frac{\partial^3 S(\rho)}{\partial\rho^3}\label{eq-qsym0},\\
Z_{\rm sym} &=& 81 \rho_0^4 \frac{\partial^4 S(\rho)}{\partial\rho^4}\label{eq-zsym0}.
\eea
The nuclear incompressibility of asymmetric matter can be elaborated in terms of isospin asymmetry $\delta = \frac{(\rho_n-\rho_p)}{\rho}$ at $\rho_0$ as $K(\delta)= K + K_{\tau} \delta^2 + \mathcal O(\delta^4)$ and $K_\tau$ is given by \cite{Chen:2009wv}
\bea
\label{ktau}
K_\tau = K_{sym} - 6 L_0 - \frac{Q_0 L_0}{K},
\eea
Where $K$ is the nuclear matter incompressibility ($9 \rho^2 \frac{\delta^2(\epsilon/\rho)} {\delta \rho^2}|_{\rho_0}$) and $Q
_0$  is the skewness parameter ($27 \rho^3 \frac{\delta^3(\epsilon/\rho)} {\delta \rho^3}|_{\rho_0}$ ). 

\section{Result and Discussion}\label{sec3}
In this section, we present the results for the density-dependence of symmetry energy, its slope, and curvature parameters over densities obtained for different cross-coupling parameters. For the present investigation, the EOS for the symmetric nuclear matter is kept unchanged by fixing the parameters $C_{\sigma}$, $C_{\omega}$, $B$ and $C$ at $\rho_0$ as obtained in Ref. \cite{Malik2017}, with the nuclear saturation properties, such as binding energy, nuclear incompressibility, and nucleon effective mass and are enlisted in Table \ref{tab1}. The values are well within the acceptable ranges prescribed in the literature. However, the effective mass $m^*/m = 0.87$ is relatively larger than some of the well-known relativistic mean field models, such as NL3 or IUFSU. It is to be noted that the magnitude of the effective mass is very important for reproducing the spin-orbital splittings of finite nuclei. If the effective nucleon mass is larger, the spin-orbital splitting will be decreased. Furthermore, the $\sigma$-meson mass is also smaller compared to other RMF models. In one of the previous works\cite{Jha:2008yth}, the correlation of nucleon effective mass and the corresponding mass of the $\sigma-$meson was analyzed. The authors found that with larger the nucleon effective mass in the model, the smaller the mass of the scalar meson to satisfy the nuclear matter saturation properties. Another rare feature of the present model is the behavior of the nucleon effective mass with density. In our case, the effective mass goes down with density but again increases after $\sim ~3\rho_0$ \cite{Jha2006, Jha:2008yth}. This behavior is attributed to the fact that repulsive forces start to dominate as the matter density increases.

\begin{table}[htp]
\caption{Model parameters for the symmetric nuclear matter calculated at  energy per nucleon ($E_0$), nuclear incompressibility ($K$), and effective nucleon mass ($m^*/m$)  at the saturation density $\rho_0$=0.153 fm$^{-3}$. $C_\sigma = g_\sigma^2/m_\sigma^2$ and $C_\omega = g_\omega^2/m_\omega^2$ are the scalar and vector meson coupling parameters, respectively. The higher-order self-couplings of scalar field parameters are $B = b/m^2$ and $C = c/m^4$, respectively, where $m$ is the mass of the nucleon. The masses of nucleon, $\omega$ meson and $\sigma$ meson are 939 MeV, 783 MeV and 444.6 MeV respectively.\label{tab1}}

 \centering
\begin{tabular}{ccccccc}
\hline\hline
   $C_{\sigma}$ & $C_{\omega}$ & $B$ &  $C$& $E_0$ & K & $m^*/m$ \\ 
   ($fm^2$) &($fm^2$) & ($fm^2$) & ($fm^4$)& (MeV)& MeV&  \\[1.5ex]
\hline 
7.221 & 1.670 & -6.435 & 0.001& -16.0& 231 &0.87 \\[1.5ex] 
\hline\hline
\end{tabular}

\end{table}

\begin{table*}[htp]
\caption{ The slope parameter of symmetry energy  for all the models is kept fixed $L_0 = 60$MeV. The values of the various symmetry energy parameters($J_0$, $J_1$,$K_{\rm sym,0}$, $Q_{\rm sym,0}$, $Z_{\rm sym,0}$, and $K_{\tau}$) as shown in Eqs. [\ref{eq-l0}-\ref{ktau}] are listed according to the model coupling schemes as discussed in the text. 
\label{tab2}}
\centering
\scalebox{1.0}{
\begin{tabular}{cccccccccc}
\hline\hline
  TPJM  &$J_1$& $J_0$& $C_{\rho}$ & $\eta_1$ &$\eta_2$  &$K_{\rm sym,0}$&$Q_{\rm sym,0}$ &$Z_{\rm sym,0}$ &$K_\tau$\\ [1.5ex]
   &(MeV) & (MeV)&   (fm$^2$) & & &(MeV) & (MeV) & (MeV) & (MeV)\\
\hline 
\multirow{3}{*}{1}&\multirow{3}{*}{23.4}& 31.5&   7.45 & -0.32 &5.20  &  -153.74&  124.97 &  703.76 &-427.04  \\
 & & 32.0&  6.78 & -0.15 & 6.99   & -191.31 &  97.22 &  2918.98 &-461.85  \\
 & & 32.5&  6.35 &-0.06 &6.48 &  -192.33 &  -37.344& 3435.22 &-482.22 \\[1.5ex]
\multirow{3}{*}{2}&\multirow{3}{*}{24.0}& 31.5&   12.56 & -0.80 &1.19   & -79.19& 302.59 &  -2209.27 &-348.94  \\
& & 32.0&  8.48 & -0.41 & 4.55  & -146.75 & 182.84 & 19.97 &-419.05 \\
& & 32.5& 7.14 & -0.17 & 6.81  &-195.20 &  125.73 &  2750.89 &-466.51 \\[1.5ex]
\multirow{3}{*}{3}& \multirow{3}{*}{24.6}& 31.5&  13.13 & -0.82 &1.05   & -76.66 & 310.06& -2624.05 &-346.07\\
&& 32.0& 13.32 & -0.79 & 1.63  &  -87.44 & 302.36 & -2343.15 &-353.06 \\
& & 32.5& 10.23 & -0.55&3.24  & -124.97 & 242.90& -1621.37 &-398.85 \\ [1.5ex]
\multirow{3}{*}{4}&\multirow{2}{*}{25.2}& 33 &12.75 & -0.68 &2.54  & -110.01& 296.53 & -2509.99 &-377.39 \\
& & 34 & 7.01 & -0.02 &8.15  &  -247.63 & 167.07&  6205.68 &-520.32 \\ [1.5ex]
 \hline\hline
\end{tabular}}
\end{table*}

\begin{figure*}[htp]
\centering
\includegraphics[height = 8 cm, width=16 cm]{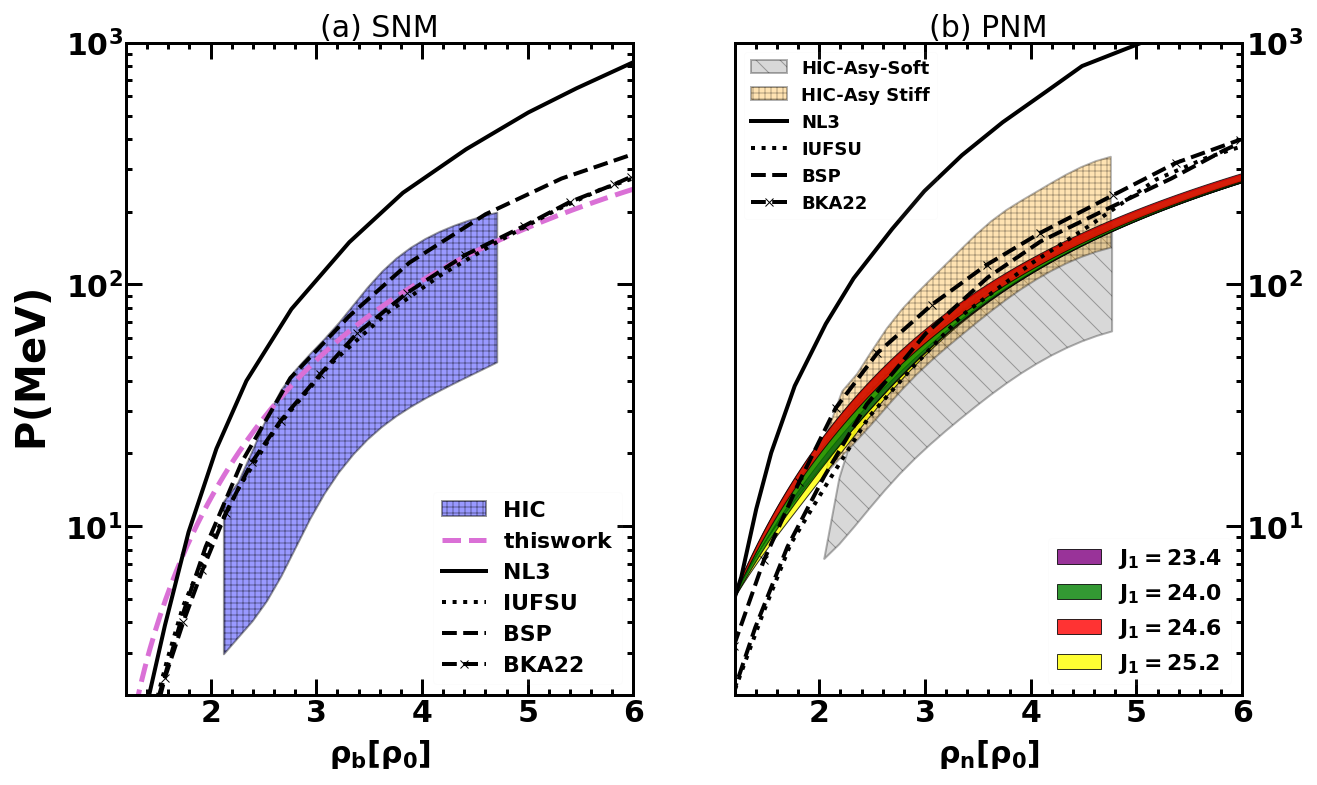}
\caption{\label{fig1}
(Color Online) The pressure is plotted against density for (left) SNM, and (right) PNM  is plotted. The results for PNM EOS  obtained for different classes of our model based on $J_1$ values are displayed and compared with a few RMF models such as NL3, IUFSU, BSP, and BKA22. The shaded regions on both panels are the constraints from Havey Ion Collision (HIC) experiments \cite{Danielewicz:2002pu}.}
\end{figure*}

For  density-dependent of symmetry energy, we have three parameters, namely $C_\rho$, $\eta_1$, and $\eta_2$. We tune these three parameters to obtain different values of $J_1$ and $J_0$, but a fixed value of slope $L_0 = 60 $ MeV, which is reasonably good for the central value obtained in different analyses \cite{Malik2017}. In Table \ref{tab2}, we present the parameter sets $C_\rho$, $\eta_1$ and $\eta_2$ along with $K_{\rm sym,0},Q_{\rm sym,0},Z_{\rm sym,0}$ for different $J_0$, $J_1$. The symmetry energy slope parameter $L_0$ for all those models is kept fixed at 60 MeV. In the following, we present the density behavior of those symmetry energy parameters (Eq.(\ref{eq-l0}$-$\ref{eq-zsym0}) ) for different $J_1$ and $J_0$, and for a fixed value of $L_0$. The value of $J_1$ in our present work lies in the range (23.4$-$25.2) MeV  obtained in Ref. \cite{Trippa:2008gr, Roca-Maza:2013mla} by analyzing the experimental data on isovector giant resonances.

The pressure for both symmetric nuclear matter (left) and pure neutron matter (right) are plotted in Fig. \ref{fig1} as a function of density. A comparison with the constraints obtained from heavy-ion collision flow (HIC) experiments and a few well-known theoretical models is shown.  It is to be noted that the HIC PNM pressure is calculated theoretically with two extreme parameterizations, the  Asy-soft (weakest) and  Asy-stiff (strongest) of symmetry energy \cite{Danielewicz:2002pu}. Here, we find that our model prediction is in good agreement with the band of the HIC data in the entire range, very similar to other models such as IUFSU \cite{Fattoyev:2010mx}, BSP \cite{Agrawal:2012rx}, and BKA22 \cite{Agrawal:2010wg}. It is to be noted that the NL3 has both a stiff SNM and a PNM content as its nuclear matter incompressibility K is 270 MeV, and the slope of the symmetry energy $L_0$ is 120 MeV at saturation density $\rho_0$. The figure also reflects the same as it is out of the HIC constraints for both SNM and PNM.  The models employed in the present study based on different classes of $J_1$ don't show any substantial difference in the pressure. 
It is interesting to note that all of our models have different symmetry energy behavior or sufficient wide dispersion in symmetry energy at high density, but it dissolves in pressure. It is because of many degeneracies of proton fraction or symmetry energy for a given pressure\cite{Imam:2021dbe, Mondal:2021vzt}.

\begin{figure}[htp]
\centering
\includegraphics[width=0.5\textwidth]{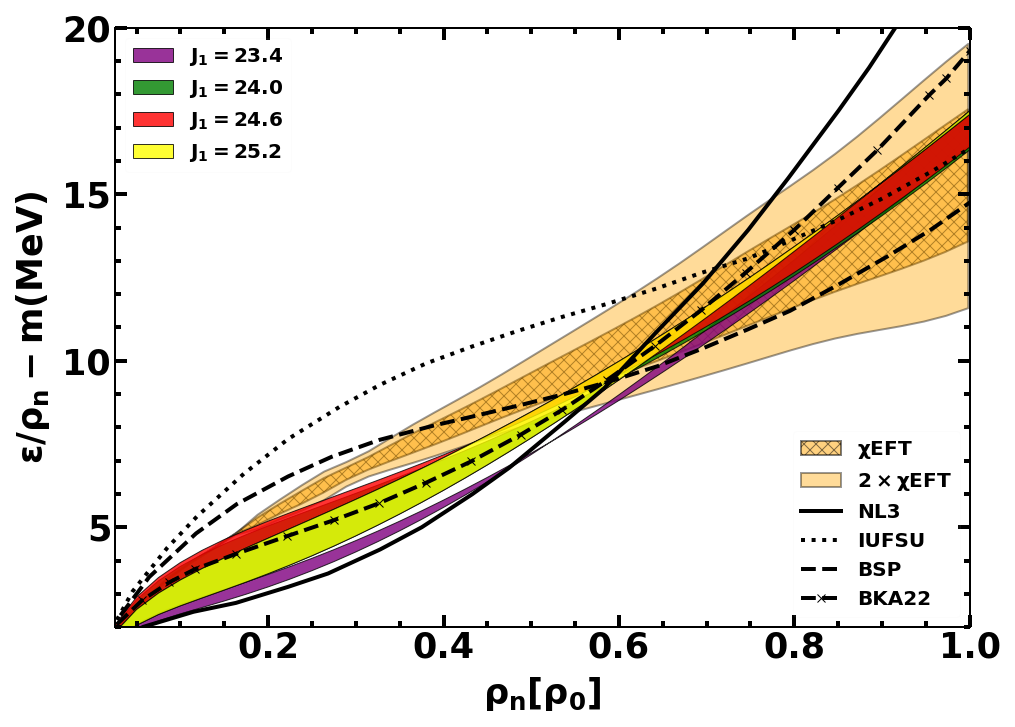}
\caption{\label{fig2}
(Color Online) Energy per neutron as a function of  density ($\rho_n$) is shown for different classes of our models based on $J_1$ values. Few selected RMF models such as NL3, IUFSU, BSP and BKA22 are also compared \cite{Gezerlis:2009iw, Hebeler:2013nza}.}
\end{figure}

\begin{figure}[hbp]
\centering
\includegraphics[width=0.5\textwidth]{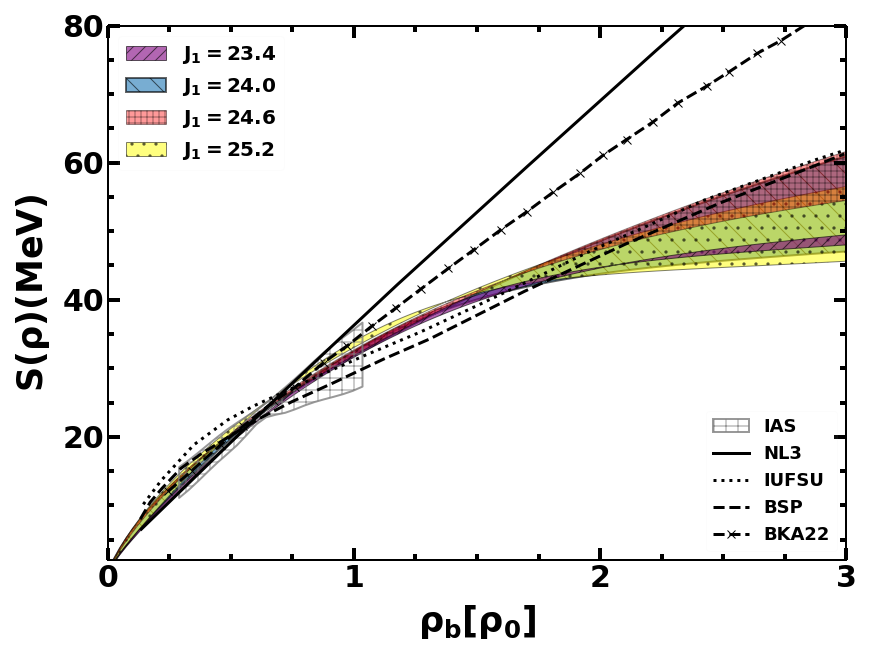}
\caption{\label{fig3}
(Color Online) Symmetry energy ($S(\rho)$) as a function of density(${\rho_b}$) is plotted for different classes of our models based on $J_1$ values along with a few selected RMF models such as NL3, IUFSU, BSP, and BKA22. The constraint of symmetry energy from the isobaric analog states (IAS) is also displayed \cite{Danielewicz:2013upa}.}
\end{figure}

 The Chiral EFT describes the hierarchy of two-, three-, and weaker higher-body forces and provides theoretical uncertainty estimates.. Microscopic calculations, which are derived from chiral effective field theory, help us to constrain the properties of neutron-rich matter up to nuclear saturation density to a high degree. In Figure \ref{fig2}, we show the energy per particle for PNM for different classes of our models based on $J_1$ values.
  We compare the pure neutron matter energy obtained from $\chi$EFT calculations in yellow hatched \cite{Hebeler:2013nza} area. We also plot this constraint with twice the uncertainty in the yellow-shaded region. As we know, this PNM constraint strongly influences symmetry energy up to supra-saturation density and is a stringent constraint for most of the RMF models. For comparison, we also show the PNM energy obtained from a few well-known RMF models, such as NL3, IUFSU, BSP, and BKA22. As we can see, our models for different $J_1$ are in agreement with the $\chi$EFT constraints from $\approx$ 0.4 $\rho_0$ density similar to others. However, the prediction of $J_1 = 24.6$ MeV is the closest to $\chi$EFT PNM constraints compared to all the models throughout the density range shown. The mesonic cross couplings, therefore, seem to be instrumental in dictating and regulating the EOS at both high and low densities, particularly with the symmetry energy aspects of the matter.
\begin{figure}[htp]
\centering
\includegraphics[width=0.5\textwidth]{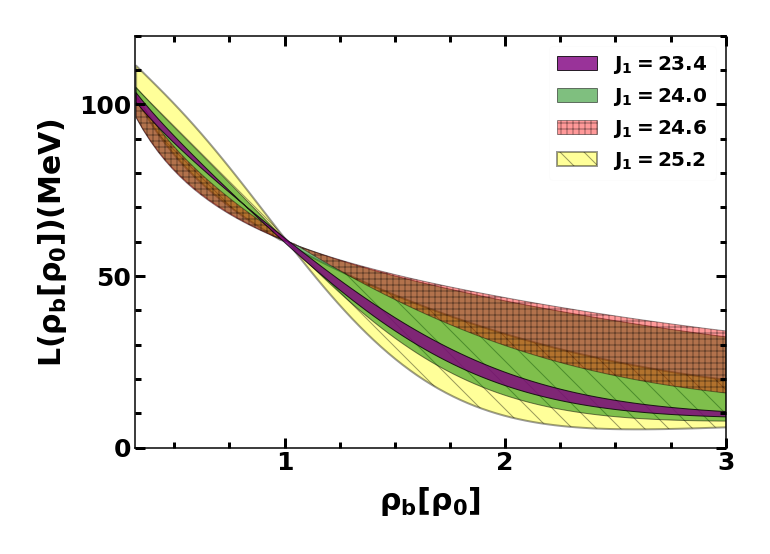}
\caption{\label{fig4}
(Color Online) The slope parameter $L$ (Eq. \ref{eq-l0}) as a function of density ($\rho_b$) is shown. Different bands of $L$ corresponding to different values of $J_1$ are displayed (see Table \ref{tab2} ).}
\end{figure}

The density content of the symmetry energy as a function of baryon density is plotted in Fig. \ref{fig3}. Additionally, we compare the symmetry energy constraint provided by the isobaric analog states (IAS) \cite{Danielewicz:2013upa}. It can be seen from the figure that all of our model's predictions agree with IAS well. We also compare the density-dependent symmetry energy derived from a few RMF models, namely NL3, IUFSU, BSP, and BKA22. All of our models for different values of $J_1$ show softer symmetry energy with a spread above two times saturation density compared to other RMF models shown in the figure. It is to be noted for all of our models, the slope of the symmetry energy at saturation is fixed to $L_0$ = 60 MeV. The values of $L_0$ are 120.65, 49.26, 50, and 78.79 for the models NL3, IUFSU, BSP, and BKA22, respectively. We find that the model with $J_1$ = 25.2 MeV represents the softest symmetry energy among others.

\begin{figure}[h!]
\centering
\includegraphics[height=10cm,width=0.5\textwidth]{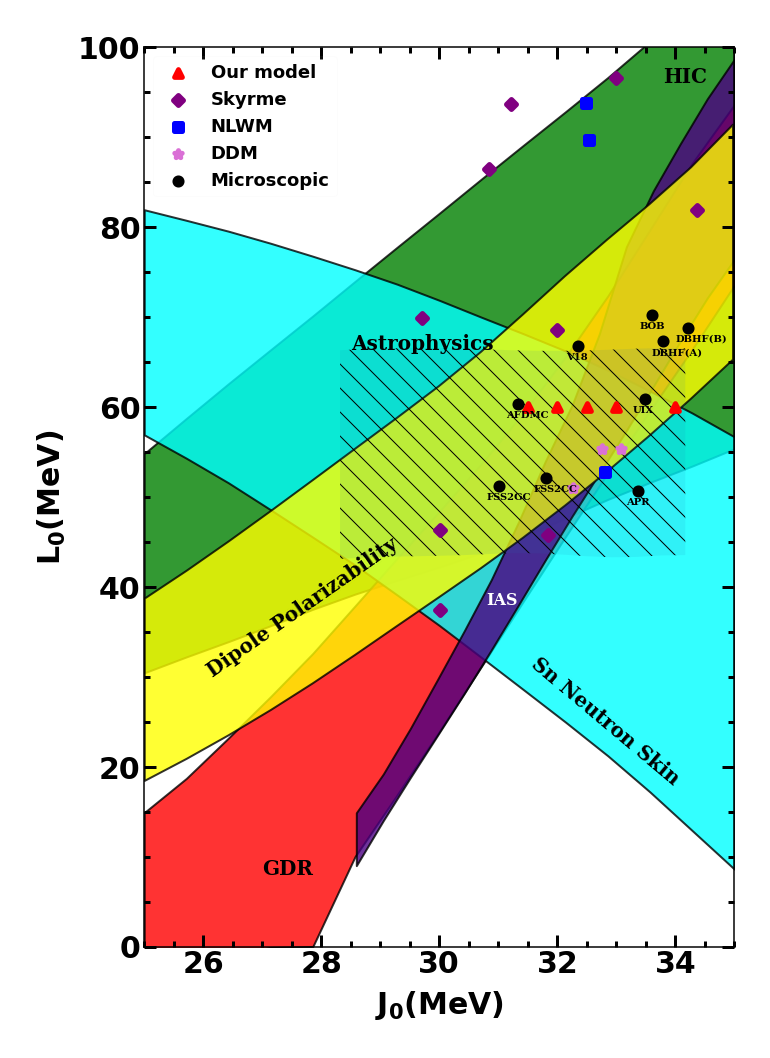}
\caption{\label{fig5}
(Color Online) The $J_0-L_0$ phase space is plotted. Present model calculations have been shown with various other theoretical approaches such as microscopic, the Skyrme, the non-linear Walecka model, and the density-dependent mesonic couplings model (DDM). The different experimental constraints such as heavy-ion collisions (HIC) \cite{Tsang:2008fd}, isobaric analog states (IAS) \cite{Danielewicz:2016bgb}, giant dipole resonances (GDR) \cite{Trippa:2008gr} those obtained from fitting astrophysical M–R observations\cite{Steiner:2010fz, Silich:2013qx}, electric dipole polarizability \cite{Roca-Maza:2015eza} and neutron-skin thicknesses of Sn isotopes \cite{Chen:2010qx} are also displayed.}
\end{figure}

In Figure \ref{fig4}, we plot the density-dependence of the slope parameter $L$ as a function of  baryon density for the models with different $J_1$ values, corresponding to the symmetry energy at saturation density, $J_0$ in the range (31.5 $-$ 34) MeV (see Table \ref{tab2}). The values agree to $J_0 \simeq 32.25\pm 2.5$ MeV as obtained from Refs. \cite{Roca-Maza:2013mla,Roca-Maza:2015eza,Vinas:2013hua,Mondal:2016roo}. The value of $L_0 = 60$ MeV is fixed for all the cases and is the median value which is obtained from recent analyses of isospin diffusion and the astrophysical data\cite{Chen:2010qx, Lattimer:2012xj, Oertel:2016bki}. Our models have a fixed value of $L_0$ at saturation density, which can be inferred from the plot, where they all converge to the same point at $\rho_0$. However, we find that the model with a lower value of $L$ below saturation density results in a step up in values after saturation density. In Fig. \ref{fig5}, We plot $J_0-L_0$ phase space for all of our models and compare them with available empirical/experimental constraints. Our model results are comparable with other theoretical models as well as various terrestrial experimental constraints and astrophysical observations. Compatibility with the overlap region of the $J_0 - L_0$ phase space is of interest as it agrees to a diverse set of constraints available. We find that our model predictions lie in this overlap region, which again speaks of its prominence in the current model.

\begin{figure}[htp]
\centering
\includegraphics[width=0.5\textwidth]{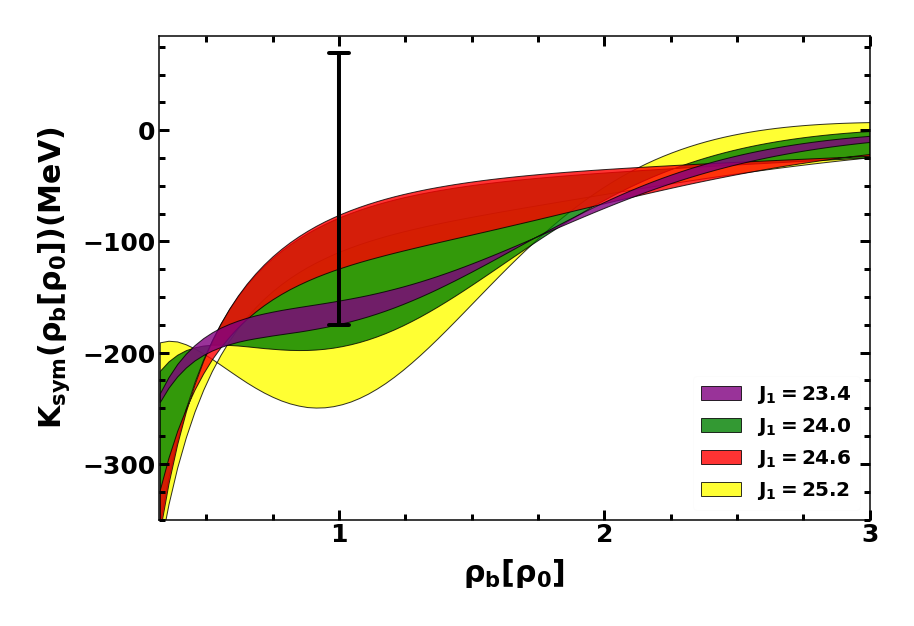}
\caption{\label{fig6}
(Color Online) The curvature parameter $K_{\rm sym}$ (Eq. \ref{eq-ksym0}) as a function of density ($\rho_b$) is shown. The results of our models  with different bands of $K_{\rm sym}$ corresponding to symmetry energy ($J_1$)  is displayed.}
\end{figure}

In Fig. \ref{fig6}, we plot the density dependence of symmetry energy curvature parameter $K_{\rm sym}$ against  baryon number density for all of our different models corresponding to different symmetry energy($J_1$). We also compare the range of $K_{\rm sym}$ at saturation density $K_{\rm sym,0} \in[-177,71]$ MeV obtained from RMF within the Bayesian Inference approach with the minimal constraint imposed \cite{Malik:2022zol, Malik:2023mnx}. All of our models show good agreement with the value at saturation density except the model with $J_1 = 25.2$ MeV. The density dependence of curvature parameter $K_{\rm sym}$, in general, shows an increasing trend over density, contrary to the behavior of $L$.

\begin{figure}[htp]
\centering
\includegraphics[width=0.5\textwidth]{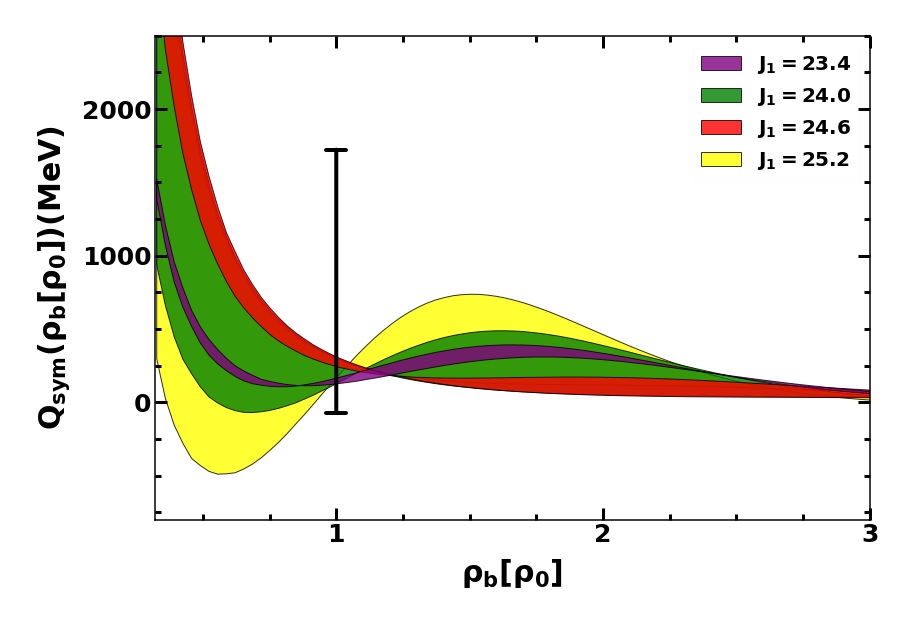}
\caption{\label{fig7}
(Color Online) The skewness parameter $Q_{\rm sym}$(Eq.\ref{eq-qsym0}) is plotted as a function of density ($\rho_b$). Different bands of $Q_{\rm sym}$ corresponding to symmetry energy $J_1$ are displayed.}
\end{figure}

\begin{figure}[htp]
\centering
\includegraphics[width=0.5\textwidth]{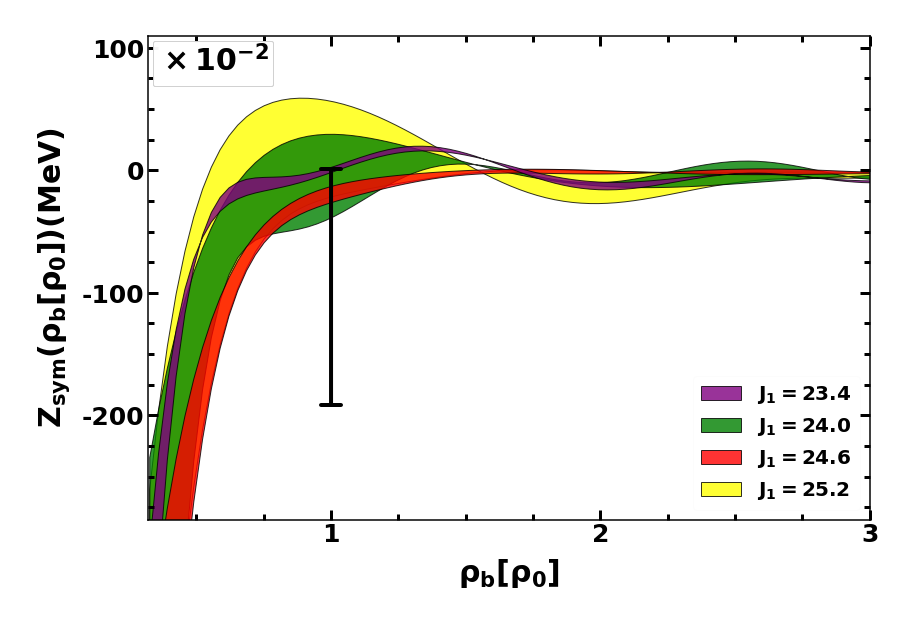}
\caption{\label{fig8}
(Color Online) The higher-order parameter $Z_{\rm sym}$ (Eq. \ref{eq-zsym0}) as a function of density ($\rho_b$). Different bands of $Z_{\rm sym}$ corresponding to symmetry energy $J_1$ are displayed.}
\end{figure}

We plot the density dependence of the symmetry energy skewness parameter $Q_{\rm sym}$ in Figure \ref{fig7} against baryon number density for all models corresponding to different symmetry energy ($J_1$). Also, we compare the range of $Q_ {\rm sym,0} \in[-88,1736]$ at saturation density
obtained from the microscopic model in RMF treatment with the minimal constraint imposed \cite{Malik:2022zol, Malik:2023mnx}. 
The values of $Q_{\rm sym,0}$ do also agree to the recently obtained range (-607.65 $-$ 783.21) MeV in Ref.\cite{dutra2014relativistic, Patra:2022yqc}. At saturation density, this parameter may have a wide range, but all of our models fall within the constraints. The density dependence of the skewness parameter $Q_{\rm sym}$ shows a decreasing trend and becomes saturated at higher densities. However, all the models show nonlinearity around saturation density except for the model with $J_1 = 24.6$ MeV. The nonlinearity is greater for models with $J_1$ equal to 25 MeV. This is because we have fixed the value of slope parameter $L_{0}$ for all models. The density dependence of the symmetry energy kurtosis parameter $Z_{\rm sym}$ against scaled baryon number density in Fig. \ref{fig8} for all models corresponding to different symmetry energy ($J_1$). We put a band of $Z_{\rm sym,0} \in[-19290,236]$ obtained in Ref \cite{Malik:2022zol, Malik:2023mnx}. Except for the model with $J_1=25.2$, all of our models show good agreement with the value at saturation density. The figure illustrates that the kurtosis parameter $Z_{\rm sym}$ increases over density and becomes saturated at a higher density above 2 $\times \rho_0$.

\begin{figure}[htp]
\centering
\includegraphics[width=0.5\textwidth]{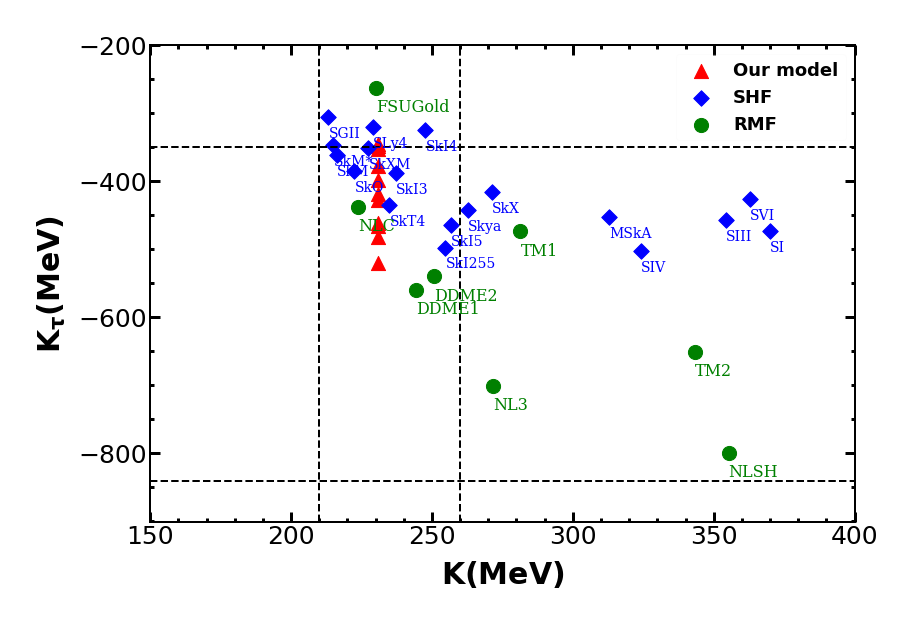}
\caption{\label{fig9}
(Color Online) The values of $K_\tau$ and $K$ of our model compared with different theoretical models \cite{Sagawa:2007sp, Colo:2013yta}. The horizontal and vertical dashed lines represent the empirical ranges of $K_\tau$ and $K$, respectively. }
\end{figure}

As mentioned earlier, the nuclear matter incompressibility $K$ and the $K_{\tau}$ are crucial in the determination of a reliable EOS as they control the stiffness of the symmetric and asymmetric parts of the EOS separately.
The influence of parameters $L_{0}$ and $K_{\rm sym,0}$ predominantly determine the value of $K_{\tau}$ and is related to the incompressibility of asymmetric nuclear matter. It is to be noted that various theoretical approaches fail to converge on a particular range of nuclear incompressibility (K), and there is also disagreement with the experimental data that is available. For example, in the non-relativistic domain, random phase calculations (RPA) calculations indicate $K$ in the range (210 - 220) MeV \cite{BLAIZOT1995, Blaizot:1980tw}, microscopic Gogny effective interactions limit the value to 231$\pm$5 MeV \cite{Youngblood:1999zza}.  Recently a range of (240$\pm$20) MeV was reported using various experimental data on isoscalar giant monopole resonances \cite{shlomo2006}. Combined together, the mean value would come to be $\sim 230$ MeV. Similarly, the constraints on $K_{\tau}$ ranges from $-840$ to $−350$ MeV \cite{Stone:2014wza,Pearson:2010zz}. In Fig. \ref{fig9}, we have shown the range on the $K$-$K_{\tau}$ phase space for our model (marked with a red triangle) along with several other well-known models and experimental constraints. 
All our models predict the incompressibility ($K$) value at 231 MeV, and we have only modified the symmetry energy component, which affects $K_\tau$ only. It can be seen that our model predictions seem to agree with the majority of the theoretical approaches. 

\begin{figure}[h!]
\centering
\includegraphics[width=0.5\textwidth]{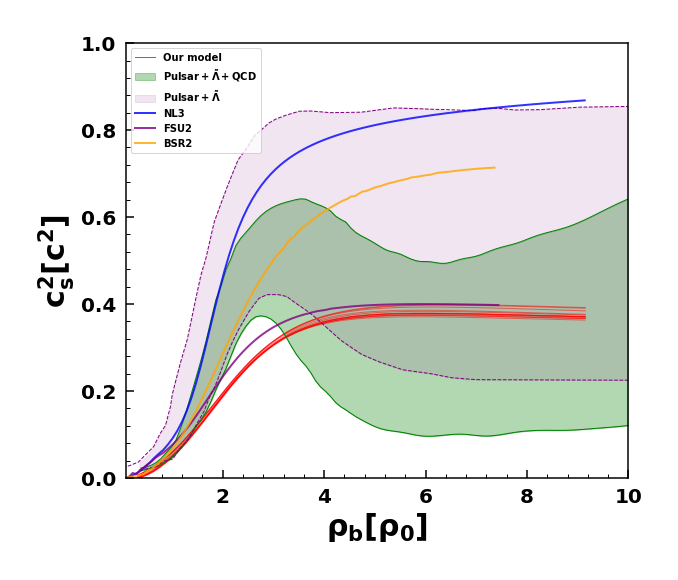}
\caption{\label{fig10}
(Color Online) The speed of sound $c_s^2$(in $c^2$) as a function of baryon density is shown. The color lines correspond to a few selected RMF models, such as NL3, BSR2, and FSU2. For comparison, we displayed the speed of the sound domain from Ref.\cite{Kurkela:2022elj} obtained with pQCD, and astrophysical constraints (shaded light green and light purple region ).}
\end{figure}

We extend our analysis to study the speed of sound of beta-equilibrated charge neutral matter inside the star. The behavior of $c_s$ at such a high density is important because it controls the stiffness of the equation of state. 
In Fig. \ref{fig10}, we plot the square of the speed of sound as a function of density for all models mentioned in Table \ref{tab2}. We observe that all of our models give a monotonically increasing speed of sound up to four times saturation density, and above this density, it gets saturated and tends to decrease towards $\approx \sqrt{1/3}$c.
 However, in all of our models, the rate of decrease is minimal (in the third decimal place only). Generally, in most microscopic models within RMF treatment, it commonly shows a monotonically increasing trend and gets saturated at a  value higher than 0.4.
 For a better comparison, we show the same obtained from other RMF models:  FSU2 \cite{chen2014building}, BSR2 \cite{Agrawal:2010wg}, and NL3 \cite{lalazissis1997new}. 
 We also compare the results obtained in Ref. \cite{Kurkela:2022elj, Gorda:2022jvk}, where the authors calculated the possible domain of speed of sound by imposing astrophysical and QCD constraints. It was shown that a knee-like feature forms at densities of approximately 3$\rho_0$, which is the critical density when QCD softens the equation of state. The speed of sound, which is obtained from the equation state, exhibits the effects of this property in the form of a peak-like structure. The speed of sound obtained with our parameter set falls in the overlap region of Astro and Astro+ QCD constraints (Mass-Radius-Tidal deformability). It can be seen from the figure that the NL3 and BSR2 do not satisfy the constraints imposed by {\it astro}+pQCD (shaded light green), whereas they are within the region obtained with {\it astro} constraints (shaded light purple) \cite{Kurkela:2022elj, Gorda:2022jvk}. It is to be noted that all of our models for different values of $J_1$ noticeably agree with those constraints.

\section{Conclusions} \label{sec4}
Studies of dense matter physics are complicated because of the unknown knowledge of nuclear forces. In dense matter studies, chiral symmetry models are promising to replicate three-body forces, which may play a vital role \cite{Lee:1974ma, Ogawa:2003ns, Chanfray:2006nz}. On the other hand, the particle content in the underlying EOS is significantly influenced by the density-dependent symmetry energy. Several phases or compositions, such as hyperons, quarks, superconducting matter, and colored super conditions, can be suitable candidates for a dense matter EOS. The value of symmetry energy at saturation density is quite known from nuclear physics experimental or empirical knowledge, however, is not well known at higher densities. In this work, we have done a case study employing the existing chiral model \cite{Malik2017} to investigate the density-dependent behavior of different slope parameters of the symmetry energy. We also investigate the behavior of the stiffness parameter of EOS: the square of the speed of sound with different values of $J_1$. 

In the present work, we bring out the essence of the density behavior of several higher-order symmetry energy parameters in the presence of various mesonic cross-couplings. At both high and low densities, mesonic cross couplings appear to be crucial in determining and controlling the EOS, especially when considering the symmetry energy features of the matter. 
For models with varying $J_1$ in the range (23.4-25.2) MeV for a fixed value of $L_0 = 60$ MeV, we found that model with $J_1 = 24.6$ shows better performance over all the empirical constraints of nuclear symmetry energy shown in this article. 
The incompressibility of nuclear matter, due to their independent influence over the symmetric and asymmetric components of the EOS, $K$, and the $K_{\tau}$, are essential in determining a reliable EOS. 

The behavior of the square of the speed of sound $c_s^2$ at such a high density is important because it controls the stiffness of the equation of state. Our model predictions for the square of the speed of sound fall within the overlap region of Astro and QCD constraints (Mass-Radius-Tidal deformability) obtained in Ref. \cite{Kurkela:2022elj, Gorda:2022jvk}. We find that up to four times the saturation density, the models predict a monotonically increasing sound speed, and thereafter, it saturates and trends toward the value of asymptotic limit $\sqrt{1/3}$c.

It is interesting to note that the square of the sound speed at high density is similar in all of our models, but the symmetry energy at high density has spread and therefore, has uncertainty in composition. In order to fully understand this, a model-independent statistical future calculation is required, which is beyond the scope of this study.

 \section{Acknowledgements} 
 N.K.P. would like to acknowledge the Department of Science and Technology (DST), India, for the support DST/INSPIRE Fellowship/2019/IF190058.
 \clearpage
\newpage

\bibliographystyle{apsrev4-1}
\bibliography{library}

\begin{thebibliography}{86}%
\makeatletter
\providecommand \@ifxundefined [1]{%
 \@ifx{#1\undefined}
}%
\providecommand \@ifnum [1]{%
 \ifnum #1\expandafter \@firstoftwo
 \else \expandafter \@secondoftwo
 \fi
}%
\providecommand \@ifx [1]{%
 \ifx #1\expandafter \@firstoftwo
 \else \expandafter \@secondoftwo
 \fi
}%
\providecommand \natexlab [1]{#1}%
\providecommand \enquote  [1]{``#1''}%
\providecommand \bibnamefont  [1]{#1}%
\providecommand \bibfnamefont [1]{#1}%
\providecommand \citenamefont [1]{#1}%
\providecommand \href@noop [0]{\@secondoftwo}%
\providecommand \href [0]{\begingroup \@sanitize@url \@href}%
\providecommand \@href[1]{\@@startlink{#1}\@@href}%
\providecommand \@@href[1]{\endgroup#1\@@endlink}%
\providecommand \@sanitize@url [0]{\catcode `\\12\catcode `\$12\catcode
  `\&12\catcode `\#12\catcode `\^12\catcode `\_12\catcode `\%12\relax}%
\providecommand \@@startlink[1]{}%
\providecommand \@@endlink[0]{}%
\providecommand \url  [0]{\begingroup\@sanitize@url \@url }%
\providecommand \@url [1]{\endgroup\@href {#1}{\urlprefix }}%
\providecommand \urlprefix  [0]{URL }%
\providecommand \Eprint [0]{\href }%
\providecommand \doibase [0]{http://dx.doi.org/}%
\providecommand \selectlanguage [0]{\@gobble}%
\providecommand \bibinfo  [0]{\@secondoftwo}%
\providecommand \bibfield  [0]{\@secondoftwo}%
\providecommand \translation [1]{[#1]}%
\providecommand \BibitemOpen [0]{}%
\providecommand \bibitemStop [0]{}%
\providecommand \bibitemNoStop [0]{.\EOS\space}%
\providecommand \EOS [0]{\spacefactor3000\relax}%
\providecommand \BibitemShut  [1]{\csname bibitem#1\endcsname}%
\let\auto@bib@innerbib\@empty
\bibitem [{\citenamefont {Lattimer}\ and\ \citenamefont
  {Prakash}(2001)}]{Lattimer:2000nx}%
  \BibitemOpen
  \bibfield  {author} {\bibinfo {author} {\bibfnamefont {J.~M.}\ \bibnamefont
  {Lattimer}}\ and\ \bibinfo {author} {\bibfnamefont {M.}~\bibnamefont
  {Prakash}},\ }\href {\doibase 10.1086/319702} {\bibfield  {journal} {\bibinfo
   {journal} {Astrophys. J.}\ }\textbf {\bibinfo {volume} {550}},\ \bibinfo
  {pages} {426} (\bibinfo {year} {2001})},\ \Eprint
  {http://arxiv.org/abs/astro-ph/0002232} {arXiv:astro-ph/0002232} \BibitemShut
  {NoStop}%
\bibitem [{\citenamefont {Watts}\ \emph {et~al.}(2016)\citenamefont {Watts}
  \emph {et~al.}}]{Watts:2016uzu}%
  \BibitemOpen
  \bibfield  {author} {\bibinfo {author} {\bibfnamefont {A.~L.}\ \bibnamefont
  {Watts}} \emph {et~al.},\ }\href {\doibase 10.1103/RevModPhys.88.021001}
  {\bibfield  {journal} {\bibinfo  {journal} {Rev. Mod. Phys.}\ }\textbf
  {\bibinfo {volume} {88}},\ \bibinfo {pages} {021001} (\bibinfo {year}
  {2016})},\ \Eprint {http://arxiv.org/abs/1602.01081} {arXiv:1602.01081
  [astro-ph.HE]} \BibitemShut {NoStop}%
\bibitem [{\citenamefont {\"Ozel}\ and\ \citenamefont
  {Freire}(2016)}]{Ozel:2016oaf}%
  \BibitemOpen
  \bibfield  {author} {\bibinfo {author} {\bibfnamefont {F.}~\bibnamefont
  {\"Ozel}}\ and\ \bibinfo {author} {\bibfnamefont {P.}~\bibnamefont
  {Freire}},\ }\href {\doibase 10.1146/annurev-astro-081915-023322} {\bibfield
  {journal} {\bibinfo  {journal} {Ann. Rev. Astron. Astrophys.}\ }\textbf
  {\bibinfo {volume} {54}},\ \bibinfo {pages} {401} (\bibinfo {year} {2016})},\
  \Eprint {http://arxiv.org/abs/1603.02698} {arXiv:1603.02698 [astro-ph.HE]}
  \BibitemShut {NoStop}%
\bibitem [{\citenamefont {Oertel}\ \emph {et~al.}(2017)\citenamefont {Oertel},
  \citenamefont {Hempel}, \citenamefont {Kl\"ahn},\ and\ \citenamefont
  {Typel}}]{Oertel:2016bki}%
  \BibitemOpen
  \bibfield  {author} {\bibinfo {author} {\bibfnamefont {M.}~\bibnamefont
  {Oertel}}, \bibinfo {author} {\bibfnamefont {M.}~\bibnamefont {Hempel}},
  \bibinfo {author} {\bibfnamefont {T.}~\bibnamefont {Kl\"ahn}}, \ and\
  \bibinfo {author} {\bibfnamefont {S.}~\bibnamefont {Typel}},\ }\href
  {\doibase 10.1103/RevModPhys.89.015007} {\bibfield  {journal} {\bibinfo
  {journal} {Rev. Mod. Phys.}\ }\textbf {\bibinfo {volume} {89}},\ \bibinfo
  {pages} {015007} (\bibinfo {year} {2017})},\ \Eprint
  {http://arxiv.org/abs/1610.03361} {arXiv:1610.03361 [astro-ph.HE]}
  \BibitemShut {NoStop}%
\bibitem [{\citenamefont {Vida\~na}(2018)}]{Vidana:2018bdi}%
  \BibitemOpen
  \bibfield  {author} {\bibinfo {author} {\bibfnamefont {I.}~\bibnamefont
  {Vida\~na}},\ }\href {\doibase 10.1098/rspa.2018.0145} {\bibfield  {journal}
  {\bibinfo  {journal} {Proc. Roy. Soc. Lond. A}\ }\textbf {\bibinfo {volume}
  {474}},\ \bibinfo {pages} {0145} (\bibinfo {year} {2018})},\ \Eprint
  {http://arxiv.org/abs/1803.00504} {arXiv:1803.00504 [nucl-th]} \BibitemShut
  {NoStop}%
\bibitem [{\citenamefont {Bombaci}\ and\ \citenamefont
  {Logoteta}(2018)}]{Bombaci:2018ksa}%
  \BibitemOpen
  \bibfield  {author} {\bibinfo {author} {\bibfnamefont {I.}~\bibnamefont
  {Bombaci}}\ and\ \bibinfo {author} {\bibfnamefont {D.}~\bibnamefont
  {Logoteta}},\ }\href {\doibase 10.1051/0004-6361/201731604} {\bibfield
  {journal} {\bibinfo  {journal} {Astron. Astrophys.}\ }\textbf {\bibinfo
  {volume} {609}},\ \bibinfo {pages} {A128} (\bibinfo {year} {2018})},\ \Eprint
  {http://arxiv.org/abs/1805.11846} {arXiv:1805.11846 [astro-ph.HE]}
  \BibitemShut {NoStop}%
\bibitem [{\citenamefont {Danielewicz}\ \emph {et~al.}(2002)\citenamefont
  {Danielewicz}, \citenamefont {Lacey},\ and\ \citenamefont
  {Lynch}}]{Danielewicz:2002pu}%
  \BibitemOpen
  \bibfield  {author} {\bibinfo {author} {\bibfnamefont {P.}~\bibnamefont
  {Danielewicz}}, \bibinfo {author} {\bibfnamefont {R.}~\bibnamefont {Lacey}},
  \ and\ \bibinfo {author} {\bibfnamefont {W.~G.}\ \bibnamefont {Lynch}},\
  }\href {\doibase 10.1126/science.1078070} {\bibfield  {journal} {\bibinfo
  {journal} {Science}\ }\textbf {\bibinfo {volume} {298}},\ \bibinfo {pages}
  {1592} (\bibinfo {year} {2002})},\ \Eprint
  {http://arxiv.org/abs/nucl-th/0208016} {arXiv:nucl-th/0208016} \BibitemShut
  {NoStop}%
\bibitem [{\citenamefont {Baran}\ \emph {et~al.}(2005)\citenamefont {Baran},
  \citenamefont {Colonna}, \citenamefont {Greco},\ and\ \citenamefont
  {Di~Toro}}]{Baran:2004ih}%
  \BibitemOpen
  \bibfield  {author} {\bibinfo {author} {\bibfnamefont {V.}~\bibnamefont
  {Baran}}, \bibinfo {author} {\bibfnamefont {M.}~\bibnamefont {Colonna}},
  \bibinfo {author} {\bibfnamefont {V.}~\bibnamefont {Greco}}, \ and\ \bibinfo
  {author} {\bibfnamefont {M.}~\bibnamefont {Di~Toro}},\ }\href {\doibase
  10.1016/j.physrep.2004.12.004} {\bibfield  {journal} {\bibinfo  {journal}
  {Phys. Rept.}\ }\textbf {\bibinfo {volume} {410}},\ \bibinfo {pages} {335}
  (\bibinfo {year} {2005})},\ \Eprint {http://arxiv.org/abs/nucl-th/0412060}
  {arXiv:nucl-th/0412060} \BibitemShut {NoStop}%
\bibitem [{\citenamefont {Li}\ \emph {et~al.}(2008)\citenamefont {Li},
  \citenamefont {Chen},\ and\ \citenamefont {Ko}}]{Li:2008gp}%
  \BibitemOpen
  \bibfield  {author} {\bibinfo {author} {\bibfnamefont {B.-A.}\ \bibnamefont
  {Li}}, \bibinfo {author} {\bibfnamefont {L.-W.}\ \bibnamefont {Chen}}, \ and\
  \bibinfo {author} {\bibfnamefont {C.~M.}\ \bibnamefont {Ko}},\ }\href
  {\doibase 10.1016/j.physrep.2008.04.005} {\bibfield  {journal} {\bibinfo
  {journal} {Phys. Rept.}\ }\textbf {\bibinfo {volume} {464}},\ \bibinfo
  {pages} {113} (\bibinfo {year} {2008})},\ \Eprint
  {http://arxiv.org/abs/0804.3580} {arXiv:0804.3580 [nucl-th]} \BibitemShut
  {NoStop}%
\bibitem [{\citenamefont {Tsang}\ \emph {et~al.}(2019)\citenamefont {Tsang},
  \citenamefont {Lynch}, \citenamefont {Danielewicz},\ and\ \citenamefont
  {Tsang}}]{Tsang:2019mlz}%
  \BibitemOpen
  \bibfield  {author} {\bibinfo {author} {\bibfnamefont {M.~B.}\ \bibnamefont
  {Tsang}}, \bibinfo {author} {\bibfnamefont {W.~G.}\ \bibnamefont {Lynch}},
  \bibinfo {author} {\bibfnamefont {P.}~\bibnamefont {Danielewicz}}, \ and\
  \bibinfo {author} {\bibfnamefont {C.~Y.}\ \bibnamefont {Tsang}},\ }\href
  {\doibase 10.1016/j.physletb.2019.06.059} {\bibfield  {journal} {\bibinfo
  {journal} {Phys. Lett. B}\ }\textbf {\bibinfo {volume} {795}},\ \bibinfo
  {pages} {533} (\bibinfo {year} {2019})},\ \Eprint
  {http://arxiv.org/abs/1906.02180} {arXiv:1906.02180 [nucl-ex]} \BibitemShut
  {NoStop}%
\bibitem [{\citenamefont {Trautmann}\ and\ \citenamefont
  {Wolter}(2012)}]{Trautmann:2012nk}%
  \BibitemOpen
  \bibfield  {author} {\bibinfo {author} {\bibfnamefont {W.}~\bibnamefont
  {Trautmann}}\ and\ \bibinfo {author} {\bibfnamefont {H.~H.}\ \bibnamefont
  {Wolter}},\ }\href {\doibase 10.1142/S0218301312300032} {\bibfield  {journal}
  {\bibinfo  {journal} {Int. J. Mod. Phys. E}\ }\textbf {\bibinfo {volume}
  {21}},\ \bibinfo {pages} {1230003} (\bibinfo {year} {2012})},\ \Eprint
  {http://arxiv.org/abs/1205.2585} {arXiv:1205.2585 [nucl-ex]} \BibitemShut
  {NoStop}%
\bibitem [{\citenamefont {L\'opez}\ and\ \citenamefont
  {Terrazas~Porras}(2017)}]{Lopez:2017ctx}%
  \BibitemOpen
  \bibfield  {author} {\bibinfo {author} {\bibfnamefont {J.~A.}\ \bibnamefont
  {L\'opez}}\ and\ \bibinfo {author} {\bibfnamefont {S.}~\bibnamefont
  {Terrazas~Porras}},\ }\href {\doibase 10.1016/j.nuclphysa.2016.09.012}
  {\bibfield  {journal} {\bibinfo  {journal} {Nucl. Phys. A}\ }\textbf
  {\bibinfo {volume} {957}},\ \bibinfo {pages} {312} (\bibinfo {year}
  {2017})}\BibitemShut {NoStop}%
\bibitem [{\citenamefont {Giuliani}\ \emph {et~al.}(2014)\citenamefont
  {Giuliani}, \citenamefont {Zheng},\ and\ \citenamefont
  {Bonasera}}]{Giuliani:2013kna}%
  \BibitemOpen
  \bibfield  {author} {\bibinfo {author} {\bibfnamefont {G.}~\bibnamefont
  {Giuliani}}, \bibinfo {author} {\bibfnamefont {H.}~\bibnamefont {Zheng}}, \
  and\ \bibinfo {author} {\bibfnamefont {A.}~\bibnamefont {Bonasera}},\ }\href
  {\doibase 10.1016/j.ppnp.2014.01.003} {\bibfield  {journal} {\bibinfo
  {journal} {Prog. Part. Nucl. Phys.}\ }\textbf {\bibinfo {volume} {76}},\
  \bibinfo {pages} {116} (\bibinfo {year} {2014})},\ \Eprint
  {http://arxiv.org/abs/1311.1811} {arXiv:1311.1811 [nucl-th]} \BibitemShut
  {NoStop}%
\bibitem [{\citenamefont {Garg}\ and\ \citenamefont
  {Col\`o}(2018)}]{Garg:2018uam}%
  \BibitemOpen
  \bibfield  {author} {\bibinfo {author} {\bibfnamefont {U.}~\bibnamefont
  {Garg}}\ and\ \bibinfo {author} {\bibfnamefont {G.}~\bibnamefont {Col\`o}},\
  }\href {\doibase 10.1016/j.ppnp.2018.03.001} {\bibfield  {journal} {\bibinfo
  {journal} {Prog. Part. Nucl. Phys.}\ }\textbf {\bibinfo {volume} {101}},\
  \bibinfo {pages} {55} (\bibinfo {year} {2018})},\ \Eprint
  {http://arxiv.org/abs/1801.03672} {arXiv:1801.03672 [nucl-ex]} \BibitemShut
  {NoStop}%
\bibitem [{\citenamefont {Ono}(2019)}]{Ono:2019jxm}%
  \BibitemOpen
  \bibfield  {author} {\bibinfo {author} {\bibfnamefont {A.}~\bibnamefont
  {Ono}},\ }\href {\doibase 10.1016/j.ppnp.2018.11.001} {\bibfield  {journal}
  {\bibinfo  {journal} {Prog. Part. Nucl. Phys.}\ }\textbf {\bibinfo {volume}
  {105}},\ \bibinfo {pages} {139} (\bibinfo {year} {2019})},\ \Eprint
  {http://arxiv.org/abs/1903.00608} {arXiv:1903.00608 [nucl-th]} \BibitemShut
  {NoStop}%
\bibitem [{\citenamefont {Roca-Maza}\ \emph {et~al.}(2013)\citenamefont
  {Roca-Maza}, \citenamefont {Centelles}, \citenamefont {Vi\~nas},
  \citenamefont {Brenna}, \citenamefont {Col\`o}, \citenamefont {Agrawal},
  \citenamefont {Paar}, \citenamefont {Piekarewicz},\ and\ \citenamefont
  {Vretenar}}]{Roca-Maza:2013mla}%
  \BibitemOpen
  \bibfield  {author} {\bibinfo {author} {\bibfnamefont {X.}~\bibnamefont
  {Roca-Maza}}, \bibinfo {author} {\bibfnamefont {M.}~\bibnamefont
  {Centelles}}, \bibinfo {author} {\bibfnamefont {X.}~\bibnamefont {Vi\~nas}},
  \bibinfo {author} {\bibfnamefont {M.}~\bibnamefont {Brenna}}, \bibinfo
  {author} {\bibfnamefont {G.}~\bibnamefont {Col\`o}}, \bibinfo {author}
  {\bibfnamefont {B.~K.}\ \bibnamefont {Agrawal}}, \bibinfo {author}
  {\bibfnamefont {N.}~\bibnamefont {Paar}}, \bibinfo {author} {\bibfnamefont
  {J.}~\bibnamefont {Piekarewicz}}, \ and\ \bibinfo {author} {\bibfnamefont
  {D.}~\bibnamefont {Vretenar}},\ }\href {\doibase 10.1103/PhysRevC.88.024316}
  {\bibfield  {journal} {\bibinfo  {journal} {Phys. Rev. C}\ }\textbf {\bibinfo
  {volume} {88}},\ \bibinfo {pages} {024316} (\bibinfo {year} {2013})},\
  \Eprint {http://arxiv.org/abs/1307.4806} {arXiv:1307.4806 [nucl-th]}
  \BibitemShut {NoStop}%
\bibitem [{\citenamefont {Roca-Maza}\ \emph {et~al.}(2015)\citenamefont
  {Roca-Maza}, \citenamefont {Vi\~nas}, \citenamefont {Centelles},
  \citenamefont {Agrawal}, \citenamefont {Colo'}, \citenamefont {Paar},
  \citenamefont {Piekarewicz},\ and\ \citenamefont
  {Vretenar}}]{Roca-Maza:2015eza}%
  \BibitemOpen
  \bibfield  {author} {\bibinfo {author} {\bibfnamefont {X.}~\bibnamefont
  {Roca-Maza}}, \bibinfo {author} {\bibfnamefont {X.}~\bibnamefont {Vi\~nas}},
  \bibinfo {author} {\bibfnamefont {M.}~\bibnamefont {Centelles}}, \bibinfo
  {author} {\bibfnamefont {B.~K.}\ \bibnamefont {Agrawal}}, \bibinfo {author}
  {\bibfnamefont {G.}~\bibnamefont {Colo'}}, \bibinfo {author} {\bibfnamefont
  {N.}~\bibnamefont {Paar}}, \bibinfo {author} {\bibfnamefont {J.}~\bibnamefont
  {Piekarewicz}}, \ and\ \bibinfo {author} {\bibfnamefont {D.}~\bibnamefont
  {Vretenar}},\ }\href {\doibase 10.1103/PhysRevC.92.064304} {\bibfield
  {journal} {\bibinfo  {journal} {Phys. Rev. C}\ }\textbf {\bibinfo {volume}
  {92}},\ \bibinfo {pages} {064304} (\bibinfo {year} {2015})},\ \Eprint
  {http://arxiv.org/abs/1510.01874} {arXiv:1510.01874 [nucl-th]} \BibitemShut
  {NoStop}%
\bibitem [{\citenamefont {Vi\~nas}\ \emph {et~al.}(2014)\citenamefont
  {Vi\~nas}, \citenamefont {Centelles}, \citenamefont {Roca-Maza},\ and\
  \citenamefont {Warda}}]{Vinas:2013hua}%
  \BibitemOpen
  \bibfield  {author} {\bibinfo {author} {\bibfnamefont {X.}~\bibnamefont
  {Vi\~nas}}, \bibinfo {author} {\bibfnamefont {M.}~\bibnamefont {Centelles}},
  \bibinfo {author} {\bibfnamefont {X.}~\bibnamefont {Roca-Maza}}, \ and\
  \bibinfo {author} {\bibfnamefont {M.}~\bibnamefont {Warda}},\ }\href
  {\doibase 10.1140/epja/i2014-14027-8} {\bibfield  {journal} {\bibinfo
  {journal} {Eur. Phys. J. A}\ }\textbf {\bibinfo {volume} {50}},\ \bibinfo
  {pages} {27} (\bibinfo {year} {2014})},\ \Eprint
  {http://arxiv.org/abs/1308.1008} {arXiv:1308.1008 [nucl-th]} \BibitemShut
  {NoStop}%
\bibitem [{\citenamefont {Mondal}\ \emph {et~al.}(2016)\citenamefont {Mondal},
  \citenamefont {Agrawal}, \citenamefont {De},\ and\ \citenamefont
  {Samaddar}}]{Mondal:2016roo}%
  \BibitemOpen
  \bibfield  {author} {\bibinfo {author} {\bibfnamefont {C.}~\bibnamefont
  {Mondal}}, \bibinfo {author} {\bibfnamefont {B.~K.}\ \bibnamefont {Agrawal}},
  \bibinfo {author} {\bibfnamefont {J.~N.}\ \bibnamefont {De}}, \ and\ \bibinfo
  {author} {\bibfnamefont {S.~K.}\ \bibnamefont {Samaddar}},\ }\href {\doibase
  10.1103/PhysRevC.93.044328} {\bibfield  {journal} {\bibinfo  {journal} {Phys.
  Rev. C}\ }\textbf {\bibinfo {volume} {93}},\ \bibinfo {pages} {044328}
  (\bibinfo {year} {2016})},\ \Eprint {http://arxiv.org/abs/1603.08645}
  {arXiv:1603.08645 [nucl-th]} \BibitemShut {NoStop}%
\bibitem [{\citenamefont {Li}\ \emph {et~al.}(2015)\citenamefont {Li},
  \citenamefont {Guo}, \citenamefont {Li}, \citenamefont {Chen}, \citenamefont
  {Fattoyev},\ and\ \citenamefont {Newton}}]{Li:2014qta}%
  \BibitemOpen
  \bibfield  {author} {\bibinfo {author} {\bibfnamefont {X.-H.}\ \bibnamefont
  {Li}}, \bibinfo {author} {\bibfnamefont {W.-J.}\ \bibnamefont {Guo}},
  \bibinfo {author} {\bibfnamefont {B.-A.}\ \bibnamefont {Li}}, \bibinfo
  {author} {\bibfnamefont {L.-W.}\ \bibnamefont {Chen}}, \bibinfo {author}
  {\bibfnamefont {F.~J.}\ \bibnamefont {Fattoyev}}, \ and\ \bibinfo {author}
  {\bibfnamefont {W.~G.}\ \bibnamefont {Newton}},\ }\href {\doibase
  10.1016/j.physletb.2015.03.005} {\bibfield  {journal} {\bibinfo  {journal}
  {Phys. Lett. B}\ }\textbf {\bibinfo {volume} {743}},\ \bibinfo {pages} {408}
  (\bibinfo {year} {2015})},\ \Eprint {http://arxiv.org/abs/1403.5577}
  {arXiv:1403.5577 [nucl-th]} \BibitemShut {NoStop}%
\bibitem [{\citenamefont {Trippa}\ \emph {et~al.}(2008)\citenamefont {Trippa},
  \citenamefont {Colo},\ and\ \citenamefont {Vigezzi}}]{Trippa:2008gr}%
  \BibitemOpen
  \bibfield  {author} {\bibinfo {author} {\bibfnamefont {L.}~\bibnamefont
  {Trippa}}, \bibinfo {author} {\bibfnamefont {G.}~\bibnamefont {Colo}}, \ and\
  \bibinfo {author} {\bibfnamefont {E.}~\bibnamefont {Vigezzi}},\ }\href
  {\doibase 10.1103/PhysRevC.77.061304} {\bibfield  {journal} {\bibinfo
  {journal} {Phys. Rev. C}\ }\textbf {\bibinfo {volume} {77}},\ \bibinfo
  {pages} {061304} (\bibinfo {year} {2008})},\ \Eprint
  {http://arxiv.org/abs/0802.3658} {arXiv:0802.3658 [nucl-th]} \BibitemShut
  {NoStop}%
\bibitem [{\citenamefont {M\"oller}\ \emph {et~al.}(2012)\citenamefont
  {M\"oller}, \citenamefont {Myers}, \citenamefont {Sagawa},\ and\
  \citenamefont {Yoshida}}]{Moller:2012pxr}%
  \BibitemOpen
  \bibfield  {author} {\bibinfo {author} {\bibfnamefont {P.}~\bibnamefont
  {M\"oller}}, \bibinfo {author} {\bibfnamefont {W.~D.}\ \bibnamefont {Myers}},
  \bibinfo {author} {\bibfnamefont {H.}~\bibnamefont {Sagawa}}, \ and\ \bibinfo
  {author} {\bibfnamefont {S.}~\bibnamefont {Yoshida}},\ }\href {\doibase
  10.1103/PhysRevLett.108.052501} {\bibfield  {journal} {\bibinfo  {journal}
  {Phys. Rev. Lett.}\ }\textbf {\bibinfo {volume} {108}},\ \bibinfo {pages}
  {052501} (\bibinfo {year} {2012})}\BibitemShut {NoStop}%
\bibitem [{\citenamefont {Tsang}\ \emph {et~al.}(2012)\citenamefont {Tsang}
  \emph {et~al.}}]{Tsang:2012se}%
  \BibitemOpen
  \bibfield  {author} {\bibinfo {author} {\bibfnamefont {M.~B.}\ \bibnamefont
  {Tsang}} \emph {et~al.},\ }\href {\doibase 10.1103/PhysRevC.86.015803}
  {\bibfield  {journal} {\bibinfo  {journal} {Phys. Rev. C}\ }\textbf {\bibinfo
  {volume} {86}},\ \bibinfo {pages} {015803} (\bibinfo {year} {2012})},\
  \Eprint {http://arxiv.org/abs/1204.0466} {arXiv:1204.0466 [nucl-ex]}
  \BibitemShut {NoStop}%
\bibitem [{\citenamefont {Lattimer}\ and\ \citenamefont
  {Lim}(2013)}]{Lattimer:2012xj}%
  \BibitemOpen
  \bibfield  {author} {\bibinfo {author} {\bibfnamefont {J.~M.}\ \bibnamefont
  {Lattimer}}\ and\ \bibinfo {author} {\bibfnamefont {Y.}~\bibnamefont {Lim}},\
  }\href {\doibase 10.1088/0004-637X/771/1/51} {\bibfield  {journal} {\bibinfo
  {journal} {Astrophys. J.}\ }\textbf {\bibinfo {volume} {771}},\ \bibinfo
  {pages} {51} (\bibinfo {year} {2013})},\ \Eprint
  {http://arxiv.org/abs/1203.4286} {arXiv:1203.4286 [nucl-th]} \BibitemShut
  {NoStop}%
\bibitem [{\citenamefont {Lattimer}(2023)}]{Lattimer:2023rpe}%
  \BibitemOpen
  \bibfield  {author} {\bibinfo {author} {\bibfnamefont {J.~M.}\ \bibnamefont
  {Lattimer}},\ }\href {\doibase 10.3390/particles6010003} {\bibfield
  {journal} {\bibinfo  {journal} {Particles}\ }\textbf {\bibinfo {volume}
  {6}},\ \bibinfo {pages} {30} (\bibinfo {year} {2023})},\ \Eprint
  {http://arxiv.org/abs/2301.03666} {arXiv:2301.03666 [nucl-th]} \BibitemShut
  {NoStop}%
\bibitem [{\citenamefont {Malik}\ \emph {et~al.}(2022)\citenamefont {Malik},
  \citenamefont {Ferreira}, \citenamefont {Agrawal},\ and\ \citenamefont
  {Provid\^encia}}]{Malik:2022zol}%
  \BibitemOpen
  \bibfield  {author} {\bibinfo {author} {\bibfnamefont {T.}~\bibnamefont
  {Malik}}, \bibinfo {author} {\bibfnamefont {M.}~\bibnamefont {Ferreira}},
  \bibinfo {author} {\bibfnamefont {B.~K.}\ \bibnamefont {Agrawal}}, \ and\
  \bibinfo {author} {\bibfnamefont {C.}~\bibnamefont {Provid\^encia}},\ }\href
  {\doibase 10.3847/1538-4357/ac5d3c} {\bibfield  {journal} {\bibinfo
  {journal} {Astrophys. J.}\ }\textbf {\bibinfo {volume} {930}},\ \bibinfo
  {pages} {17} (\bibinfo {year} {2022})},\ \Eprint
  {http://arxiv.org/abs/2201.12552} {arXiv:2201.12552 [nucl-th]} \BibitemShut
  {NoStop}%
\bibitem [{\citenamefont {Walecka}(1974)}]{Walecka:1974qa}%
  \BibitemOpen
  \bibfield  {author} {\bibinfo {author} {\bibfnamefont {J.~D.}\ \bibnamefont
  {Walecka}},\ }\href {\doibase 10.1016/0003-4916(74)90208-5} {\bibfield
  {journal} {\bibinfo  {journal} {Annals Phys.}\ }\textbf {\bibinfo {volume}
  {83}},\ \bibinfo {pages} {491} (\bibinfo {year} {1974})}\BibitemShut
  {NoStop}%
\bibitem [{\citenamefont {Serot}\ and\ \citenamefont
  {Walecka}(1986)}]{Serot1986}%
  \BibitemOpen
  \bibfield  {author} {\bibinfo {author} {\bibfnamefont {B.~D.}\ \bibnamefont
  {Serot}}\ and\ \bibinfo {author} {\bibfnamefont {J.~D.}\ \bibnamefont
  {Walecka}},\ }\href@noop {} {\bibfield  {journal} {\bibinfo  {journal} {Adv.
  Nucl. Phys.}\ }\textbf {\bibinfo {volume} {16}},\ \bibinfo {pages} {1}
  (\bibinfo {year} {1986})}\BibitemShut {NoStop}%
\bibitem [{\citenamefont {Gambhir}\ \emph {et~al.}(1990)\citenamefont
  {Gambhir}, \citenamefont {Ring},\ and\ \citenamefont
  {Thimet}}]{Gambhir:1990uyn}%
  \BibitemOpen
  \bibfield  {author} {\bibinfo {author} {\bibfnamefont {Y.~K.}\ \bibnamefont
  {Gambhir}}, \bibinfo {author} {\bibfnamefont {P.}~\bibnamefont {Ring}}, \
  and\ \bibinfo {author} {\bibfnamefont {A.}~\bibnamefont {Thimet}},\ }\href
  {\doibase 10.1016/0003-4916(90)90330-Q} {\bibfield  {journal} {\bibinfo
  {journal} {Annals Phys.}\ }\textbf {\bibinfo {volume} {198}},\ \bibinfo
  {pages} {132} (\bibinfo {year} {1990})}\BibitemShut {NoStop}%
\bibitem [{\citenamefont {Patra}\ and\ \citenamefont
  {Praharaj}(1991)}]{Patra:1991wy}%
  \BibitemOpen
  \bibfield  {author} {\bibinfo {author} {\bibfnamefont {S.~K.}\ \bibnamefont
  {Patra}}\ and\ \bibinfo {author} {\bibfnamefont {C.~R.}\ \bibnamefont
  {Praharaj}},\ }\href {\doibase 10.1103/PhysRevC.44.2552} {\bibfield
  {journal} {\bibinfo  {journal} {Phys. Rev. C}\ }\textbf {\bibinfo {volume}
  {44}},\ \bibinfo {pages} {2552} (\bibinfo {year} {1991})}\BibitemShut
  {NoStop}%
\bibitem [{\citenamefont {Serot}(2002)}]{Serot:2002ei}%
  \BibitemOpen
  \bibfield  {author} {\bibinfo {author} {\bibfnamefont {B.~D.}\ \bibnamefont
  {Serot}},\ }\href {\doibase 10.1142/S0217979203020284} {\bibfield  {journal}
  {\bibinfo  {journal} {Ser. Adv. Quant. Many Body Theor.}\ }\textbf {\bibinfo
  {volume} {6}},\ \bibinfo {pages} {207} (\bibinfo {year} {2002})},\ \Eprint
  {http://arxiv.org/abs/nucl-th/0201083} {arXiv:nucl-th/0201083} \BibitemShut
  {NoStop}%
\bibitem [{\citenamefont {Todd-Rutel}\ and\ \citenamefont
  {Piekarewicz}(2005)}]{Todd-Rutel:2005yzo}%
  \BibitemOpen
  \bibfield  {author} {\bibinfo {author} {\bibfnamefont {B.~G.}\ \bibnamefont
  {Todd-Rutel}}\ and\ \bibinfo {author} {\bibfnamefont {J.}~\bibnamefont
  {Piekarewicz}},\ }\href {\doibase 10.1103/PhysRevLett.95.122501} {\bibfield
  {journal} {\bibinfo  {journal} {Phys. Rev. Lett.}\ }\textbf {\bibinfo
  {volume} {95}},\ \bibinfo {pages} {122501} (\bibinfo {year} {2005})},\
  \Eprint {http://arxiv.org/abs/nucl-th/0504034} {arXiv:nucl-th/0504034}
  \BibitemShut {NoStop}%
\bibitem [{\citenamefont {Agrawal}(2010)}]{Agrawal:2010wg}%
  \BibitemOpen
  \bibfield  {author} {\bibinfo {author} {\bibfnamefont {B.~K.}\ \bibnamefont
  {Agrawal}},\ }\href {\doibase 10.1103/PhysRevC.81.034323} {\bibfield
  {journal} {\bibinfo  {journal} {Phys. Rev. C}\ }\textbf {\bibinfo {volume}
  {81}},\ \bibinfo {pages} {034323} (\bibinfo {year} {2010})},\ \Eprint
  {http://arxiv.org/abs/1003.3295} {arXiv:1003.3295 [nucl-th]} \BibitemShut
  {NoStop}%
\bibitem [{\citenamefont {Kumar}\ \emph {et~al.}(2017)\citenamefont {Kumar},
  \citenamefont {Singh}, \citenamefont {Agrawal},\ and\ \citenamefont
  {Patra}}]{Kumar:2017xdy}%
  \BibitemOpen
  \bibfield  {author} {\bibinfo {author} {\bibfnamefont {B.}~\bibnamefont
  {Kumar}}, \bibinfo {author} {\bibfnamefont {S.~K.}\ \bibnamefont {Singh}},
  \bibinfo {author} {\bibfnamefont {B.~K.}\ \bibnamefont {Agrawal}}, \ and\
  \bibinfo {author} {\bibfnamefont {S.~K.}\ \bibnamefont {Patra}},\ }\href
  {\doibase 10.1016/j.nuclphysa.2017.07.001} {\bibfield  {journal} {\bibinfo
  {journal} {Nucl. Phys. A}\ }\textbf {\bibinfo {volume} {966}},\ \bibinfo
  {pages} {197} (\bibinfo {year} {2017})},\ \Eprint
  {http://arxiv.org/abs/1705.02621} {arXiv:1705.02621 [nucl-th]} \BibitemShut
  {NoStop}%
\bibitem [{\citenamefont {Gell-Mann}\ and\ \citenamefont
  {Levy}(1960)}]{Gell-Mann:1960mvl}%
  \BibitemOpen
  \bibfield  {author} {\bibinfo {author} {\bibfnamefont {M.}~\bibnamefont
  {Gell-Mann}}\ and\ \bibinfo {author} {\bibfnamefont {M.}~\bibnamefont
  {Levy}},\ }\href {\doibase 10.1007/BF02859738} {\bibfield  {journal}
  {\bibinfo  {journal} {Nuovo Cim.}\ }\textbf {\bibinfo {volume} {16}},\
  \bibinfo {pages} {705} (\bibinfo {year} {1960})}\BibitemShut {NoStop}%
\bibitem [{\citenamefont {Lee}\ and\ \citenamefont {Wick}(1974)}]{Lee:1974ma}%
  \BibitemOpen
  \bibfield  {author} {\bibinfo {author} {\bibfnamefont {T.~D.}\ \bibnamefont
  {Lee}}\ and\ \bibinfo {author} {\bibfnamefont {G.~C.}\ \bibnamefont {Wick}},\
  }\href {\doibase 10.1103/PhysRevD.9.2291} {\bibfield  {journal} {\bibinfo
  {journal} {Phys. Rev. D}\ }\textbf {\bibinfo {volume} {9}},\ \bibinfo {pages}
  {2291} (\bibinfo {year} {1974})}\BibitemShut {NoStop}%
\bibitem [{\citenamefont {Ogawa}\ \emph {et~al.}(2004)\citenamefont {Ogawa},
  \citenamefont {Toki}, \citenamefont {Tamenaga}, \citenamefont {Shen},
  \citenamefont {Hosaka}, \citenamefont {Sugimoto},\ and\ \citenamefont
  {Ikeda}}]{Ogawa:2003ns}%
  \BibitemOpen
  \bibfield  {author} {\bibinfo {author} {\bibfnamefont {Y.}~\bibnamefont
  {Ogawa}}, \bibinfo {author} {\bibfnamefont {H.}~\bibnamefont {Toki}},
  \bibinfo {author} {\bibfnamefont {S.}~\bibnamefont {Tamenaga}}, \bibinfo
  {author} {\bibfnamefont {H.}~\bibnamefont {Shen}}, \bibinfo {author}
  {\bibfnamefont {A.}~\bibnamefont {Hosaka}}, \bibinfo {author} {\bibfnamefont
  {S.}~\bibnamefont {Sugimoto}}, \ and\ \bibinfo {author} {\bibfnamefont
  {K.}~\bibnamefont {Ikeda}},\ }\href {\doibase 10.1143/PTP.111.75} {\bibfield
  {journal} {\bibinfo  {journal} {Prog. Theor. Phys.}\ }\textbf {\bibinfo
  {volume} {111}},\ \bibinfo {pages} {75} (\bibinfo {year} {2004})},\ \Eprint
  {http://arxiv.org/abs/nucl-th/0312042} {arXiv:nucl-th/0312042} \BibitemShut
  {NoStop}%
\bibitem [{\citenamefont {Chanfray}\ and\ \citenamefont
  {Ericson}(2007)}]{Chanfray:2006nz}%
  \BibitemOpen
  \bibfield  {author} {\bibinfo {author} {\bibfnamefont {G.}~\bibnamefont
  {Chanfray}}\ and\ \bibinfo {author} {\bibfnamefont {M.}~\bibnamefont
  {Ericson}},\ }\href {\doibase 10.1103/PhysRevC.75.015206} {\bibfield
  {journal} {\bibinfo  {journal} {Phys. Rev. C}\ }\textbf {\bibinfo {volume}
  {75}},\ \bibinfo {pages} {015206} (\bibinfo {year} {2007})},\ \Eprint
  {http://arxiv.org/abs/nucl-th/0611042} {arXiv:nucl-th/0611042} \BibitemShut
  {NoStop}%
\bibitem [{\citenamefont {Thomas}(1984)}]{Thomas:1982kv}%
  \BibitemOpen
  \bibfield  {author} {\bibinfo {author} {\bibfnamefont {A.~W.}\ \bibnamefont
  {Thomas}},\ }\href {\doibase 10.1007/978-1-4613-9892-9_1} {\bibfield
  {journal} {\bibinfo  {journal} {Adv. Nucl. Phys.}\ }\textbf {\bibinfo
  {volume} {13}},\ \bibinfo {pages} {1} (\bibinfo {year} {1984})}\BibitemShut
  {NoStop}%
\bibitem [{\citenamefont {Lee}\ and\ \citenamefont
  {Margulies}(1975)}]{Lee:1974uu}%
  \BibitemOpen
  \bibfield  {author} {\bibinfo {author} {\bibfnamefont {T.~D.}\ \bibnamefont
  {Lee}}\ and\ \bibinfo {author} {\bibfnamefont {M.}~\bibnamefont
  {Margulies}},\ }\href {\doibase 10.1103/PhysRevD.12.4008} {\bibfield
  {journal} {\bibinfo  {journal} {Phys. Rev. D}\ }\textbf {\bibinfo {volume}
  {11}},\ \bibinfo {pages} {1591} (\bibinfo {year} {1975})}\BibitemShut
  {NoStop}%
\bibitem [{\citenamefont {Thomas}\ \emph {et~al.}(2004)\citenamefont {Thomas},
  \citenamefont {Guichon}, \citenamefont {Leinweber},\ and\ \citenamefont
  {Young}}]{Thomas:2004iw}%
  \BibitemOpen
  \bibfield  {author} {\bibinfo {author} {\bibfnamefont {A.~W.}\ \bibnamefont
  {Thomas}}, \bibinfo {author} {\bibfnamefont {P.~A.~M.}\ \bibnamefont
  {Guichon}}, \bibinfo {author} {\bibfnamefont {D.~B.}\ \bibnamefont
  {Leinweber}}, \ and\ \bibinfo {author} {\bibfnamefont {R.~D.}\ \bibnamefont
  {Young}},\ }\href {\doibase 10.1143/PTPS.156.124} {\bibfield  {journal}
  {\bibinfo  {journal} {Prog. Theor. Phys. Suppl.}\ }\textbf {\bibinfo {volume}
  {156}},\ \bibinfo {pages} {124} (\bibinfo {year} {2004})},\ \Eprint
  {http://arxiv.org/abs/nucl-th/0411014} {arXiv:nucl-th/0411014} \BibitemShut
  {NoStop}%
\bibitem [{\citenamefont {Furnstahl}\ and\ \citenamefont
  {Serot}(1993)}]{Furnstahl:1993wx}%
  \BibitemOpen
  \bibfield  {author} {\bibinfo {author} {\bibfnamefont {R.~J.}\ \bibnamefont
  {Furnstahl}}\ and\ \bibinfo {author} {\bibfnamefont {B.~D.}\ \bibnamefont
  {Serot}},\ }\href {\doibase 10.1016/0370-2693(93)90649-3} {\bibfield
  {journal} {\bibinfo  {journal} {Phys. Lett. B}\ }\textbf {\bibinfo {volume}
  {316}},\ \bibinfo {pages} {12} (\bibinfo {year} {1993})}\BibitemShut
  {NoStop}%
\bibitem [{\citenamefont {Heide}\ \emph {et~al.}(1994)\citenamefont {Heide},
  \citenamefont {Rudaz},\ and\ \citenamefont {Ellis}}]{Heide:1993yz}%
  \BibitemOpen
  \bibfield  {author} {\bibinfo {author} {\bibfnamefont {E.~K.}\ \bibnamefont
  {Heide}}, \bibinfo {author} {\bibfnamefont {S.}~\bibnamefont {Rudaz}}, \ and\
  \bibinfo {author} {\bibfnamefont {P.~J.}\ \bibnamefont {Ellis}},\ }\href
  {\doibase 10.1016/0375-9474(94)90717-X} {\bibfield  {journal} {\bibinfo
  {journal} {Nucl. Phys. A}\ }\textbf {\bibinfo {volume} {571}},\ \bibinfo
  {pages} {713} (\bibinfo {year} {1994})},\ \Eprint
  {http://arxiv.org/abs/nucl-th/9308002} {arXiv:nucl-th/9308002} \BibitemShut
  {NoStop}%
\bibitem [{\citenamefont {Mishustin}\ \emph {et~al.}(1993)\citenamefont
  {Mishustin}, \citenamefont {Bondorf},\ and\ \citenamefont
  {Rho}}]{Mishustin:1993ub}%
  \BibitemOpen
  \bibfield  {author} {\bibinfo {author} {\bibfnamefont {I.}~\bibnamefont
  {Mishustin}}, \bibinfo {author} {\bibfnamefont {J.}~\bibnamefont {Bondorf}},
  \ and\ \bibinfo {author} {\bibfnamefont {M.}~\bibnamefont {Rho}},\ }\href
  {\doibase 10.1016/0375-9474(93)90319-S} {\bibfield  {journal} {\bibinfo
  {journal} {Nucl. Phys. A}\ }\textbf {\bibinfo {volume} {555}},\ \bibinfo
  {pages} {215} (\bibinfo {year} {1993})}\BibitemShut {NoStop}%
\bibitem [{\citenamefont {Papazoglou}\ \emph {et~al.}(1999)\citenamefont
  {Papazoglou}, \citenamefont {Zschiesche}, \citenamefont {Schramm},
  \citenamefont {Schaffner-Bielich}, \citenamefont {Stoecker},\ and\
  \citenamefont {Greiner}}]{Papazoglou:1998vr}%
  \BibitemOpen
  \bibfield  {author} {\bibinfo {author} {\bibfnamefont {P.}~\bibnamefont
  {Papazoglou}}, \bibinfo {author} {\bibfnamefont {D.}~\bibnamefont
  {Zschiesche}}, \bibinfo {author} {\bibfnamefont {S.}~\bibnamefont {Schramm}},
  \bibinfo {author} {\bibfnamefont {J.}~\bibnamefont {Schaffner-Bielich}},
  \bibinfo {author} {\bibfnamefont {H.}~\bibnamefont {Stoecker}}, \ and\
  \bibinfo {author} {\bibfnamefont {W.}~\bibnamefont {Greiner}},\ }\href
  {\doibase 10.1103/PhysRevC.59.411} {\bibfield  {journal} {\bibinfo  {journal}
  {Phys. Rev. C}\ }\textbf {\bibinfo {volume} {59}},\ \bibinfo {pages} {411}
  (\bibinfo {year} {1999})},\ \Eprint {http://arxiv.org/abs/nucl-th/9806087}
  {arXiv:nucl-th/9806087} \BibitemShut {NoStop}%
\bibitem [{\citenamefont {Schramm}(2002)}]{Schramm:2002xi}%
  \BibitemOpen
  \bibfield  {author} {\bibinfo {author} {\bibfnamefont {S.}~\bibnamefont
  {Schramm}},\ }\href {\doibase 10.1103/PhysRevC.66.064310} {\bibfield
  {journal} {\bibinfo  {journal} {Phys. Rev. C}\ }\textbf {\bibinfo {volume}
  {66}},\ \bibinfo {pages} {064310} (\bibinfo {year} {2002})},\ \Eprint
  {http://arxiv.org/abs/nucl-th/0207060} {arXiv:nucl-th/0207060} \BibitemShut
  {NoStop}%
\bibitem [{\citenamefont {Tsubakihara}\ and\ \citenamefont
  {Ohnishi}(2007)}]{Tsubakihara:2006se}%
  \BibitemOpen
  \bibfield  {author} {\bibinfo {author} {\bibfnamefont {K.}~\bibnamefont
  {Tsubakihara}}\ and\ \bibinfo {author} {\bibfnamefont {A.}~\bibnamefont
  {Ohnishi}},\ }\href {\doibase 10.1143/PTP.117.903} {\bibfield  {journal}
  {\bibinfo  {journal} {Prog. Theor. Phys.}\ }\textbf {\bibinfo {volume}
  {117}},\ \bibinfo {pages} {903} (\bibinfo {year} {2007})},\ \Eprint
  {http://arxiv.org/abs/nucl-th/0607046} {arXiv:nucl-th/0607046} \BibitemShut
  {NoStop}%
\bibitem [{\citenamefont {Tsubakihara}\ \emph {et~al.}(2010)\citenamefont
  {Tsubakihara}, \citenamefont {Maekawa}, \citenamefont {Matsumiya},\ and\
  \citenamefont {Ohnishi}}]{Tsubakihara:2009zb}%
  \BibitemOpen
  \bibfield  {author} {\bibinfo {author} {\bibfnamefont {K.}~\bibnamefont
  {Tsubakihara}}, \bibinfo {author} {\bibfnamefont {H.}~\bibnamefont
  {Maekawa}}, \bibinfo {author} {\bibfnamefont {H.}~\bibnamefont {Matsumiya}},
  \ and\ \bibinfo {author} {\bibfnamefont {A.}~\bibnamefont {Ohnishi}},\ }\href
  {\doibase 10.1103/PhysRevC.81.065206} {\bibfield  {journal} {\bibinfo
  {journal} {Phys. Rev. C}\ }\textbf {\bibinfo {volume} {81}},\ \bibinfo
  {pages} {065206} (\bibinfo {year} {2010})},\ \Eprint
  {http://arxiv.org/abs/0909.5058} {arXiv:0909.5058 [nucl-th]} \BibitemShut
  {NoStop}%
\bibitem [{\citenamefont {Boguta}(1983)}]{Boguta:1983uz}%
  \BibitemOpen
  \bibfield  {author} {\bibinfo {author} {\bibfnamefont {J.}~\bibnamefont
  {Boguta}},\ }\href {\doibase 10.1016/0370-2693(83)90065-5} {\bibfield
  {journal} {\bibinfo  {journal} {Phys. Lett. B}\ }\textbf {\bibinfo {volume}
  {128}},\ \bibinfo {pages} {19} (\bibinfo {year} {1983})}\BibitemShut
  {NoStop}%
\bibitem [{\citenamefont {Sahu}\ \emph {et~al.}(1993)\citenamefont {Sahu},
  \citenamefont {Basu},\ and\ \citenamefont {Datta}}]{Sahu:1993db}%
  \BibitemOpen
  \bibfield  {author} {\bibinfo {author} {\bibfnamefont {P.~K.}\ \bibnamefont
  {Sahu}}, \bibinfo {author} {\bibfnamefont {R.}~\bibnamefont {Basu}}, \ and\
  \bibinfo {author} {\bibfnamefont {B.}~\bibnamefont {Datta}},\ }\href
  {\doibase 10.1086/173233} {\bibfield  {journal} {\bibinfo  {journal}
  {Astrophys. J.}\ }\textbf {\bibinfo {volume} {416}},\ \bibinfo {pages} {267}
  (\bibinfo {year} {1993})}\BibitemShut {NoStop}%
\bibitem [{\citenamefont {Sahu}\ \emph {et~al.}(2004)\citenamefont {Sahu},
  \citenamefont {Jha}, \citenamefont {Panda},\ and\ \citenamefont
  {Patra}}]{Sahu2004}%
  \BibitemOpen
  \bibfield  {author} {\bibinfo {author} {\bibfnamefont {P.~K.}\ \bibnamefont
  {Sahu}}, \bibinfo {author} {\bibfnamefont {T.~K.}\ \bibnamefont {Jha}},
  \bibinfo {author} {\bibfnamefont {K.~C.}\ \bibnamefont {Panda}}, \ and\
  \bibinfo {author} {\bibfnamefont {S.~K.}\ \bibnamefont {Patra}},\ }\href@noop
  {} {\bibfield  {journal} {\bibinfo  {journal} {Nucl. Phys.}\ }\textbf
  {\bibinfo {volume} {A733}},\ \bibinfo {pages} {169} (\bibinfo {year}
  {2004})}\BibitemShut {NoStop}%
\bibitem [{\citenamefont {Jha}\ and\ \citenamefont
  {Mishra}(2008)}]{Jha:2008yth}%
  \BibitemOpen
  \bibfield  {author} {\bibinfo {author} {\bibfnamefont {T.~K.}\ \bibnamefont
  {Jha}}\ and\ \bibinfo {author} {\bibfnamefont {H.}~\bibnamefont {Mishra}},\
  }\href {\doibase 10.1103/PhysRevC.78.065802} {\bibfield  {journal} {\bibinfo
  {journal} {Phys. Rev. C}\ }\textbf {\bibinfo {volume} {78}},\ \bibinfo
  {pages} {065802} (\bibinfo {year} {2008})},\ \Eprint
  {http://arxiv.org/abs/0811.4233} {arXiv:0811.4233 [nucl-th]} \BibitemShut
  {NoStop}%
\bibitem [{\citenamefont {Jha}\ \emph {et~al.}(2006)\citenamefont {Jha},
  \citenamefont {Raina}, \citenamefont {Panda},\ and\ \citenamefont
  {Patra}}]{Jha2006}%
  \BibitemOpen
  \bibfield  {author} {\bibinfo {author} {\bibfnamefont {T.~K.}\ \bibnamefont
  {Jha}}, \bibinfo {author} {\bibfnamefont {P.~K.}\ \bibnamefont {Raina}},
  \bibinfo {author} {\bibfnamefont {P.~K.}\ \bibnamefont {Panda}}, \ and\
  \bibinfo {author} {\bibfnamefont {S.~K.}\ \bibnamefont {Patra}},\ }\href
  {\doibase 10.1103/PhysRevC.74.055803, 10.1103/PhysRevC.75.029903} {\bibfield
  {journal} {\bibinfo  {journal} {Phys. Rev. C}\ }\textbf {\bibinfo {volume}
  {74}},\ \bibinfo {pages} {055803} (\bibinfo {year} {2006})},\ \Eprint
  {http://arxiv.org/abs/nucl-th/0608013} {arXiv:nucl-th/0608013 [nucl-th]}
  \BibitemShut {NoStop}%
\bibitem [{\citenamefont {Jha}\ \emph {et~al.}(2010)\citenamefont {Jha},
  \citenamefont {Mishra},\ and\ \citenamefont {Sreekanth}}]{Jha2010}%
  \BibitemOpen
  \bibfield  {author} {\bibinfo {author} {\bibfnamefont {T.~K.}\ \bibnamefont
  {Jha}}, \bibinfo {author} {\bibfnamefont {H.}~\bibnamefont {Mishra}}, \ and\
  \bibinfo {author} {\bibfnamefont {V.}~\bibnamefont {Sreekanth}},\ }\href
  {\doibase 10.1103/PhysRevC.82.025803} {\bibfield  {journal} {\bibinfo
  {journal} {Phys. Rev. C}\ }\textbf {\bibinfo {volume} {82}},\ \bibinfo
  {pages} {025803} (\bibinfo {year} {2010})}\BibitemShut {NoStop}%
\bibitem [{\citenamefont {Malik}\ \emph {et~al.}(2017)\citenamefont {Malik},
  \citenamefont {Banerjee}, \citenamefont {Jha},\ and\ \citenamefont
  {Agrawal}}]{Malik2017}%
  \BibitemOpen
  \bibfield  {author} {\bibinfo {author} {\bibfnamefont {T.}~\bibnamefont
  {Malik}}, \bibinfo {author} {\bibfnamefont {K.}~\bibnamefont {Banerjee}},
  \bibinfo {author} {\bibfnamefont {T.~K.}\ \bibnamefont {Jha}}, \ and\
  \bibinfo {author} {\bibfnamefont {B.~K.}\ \bibnamefont {Agrawal}},\ }\href
  {\doibase 10.1103/PhysRevC.96.035803} {\bibfield  {journal} {\bibinfo
  {journal} {Phys. Rev.}\ }\textbf {\bibinfo {volume} {C96}},\ \bibinfo {pages}
  {035803} (\bibinfo {year} {2017})},\ \Eprint
  {http://arxiv.org/abs/1708.07291} {arXiv:1708.07291 [nucl-th]} \BibitemShut
  {NoStop}%
\bibitem [{\citenamefont {Patra}\ \emph {et~al.}(2020)\citenamefont {Patra},
  \citenamefont {Malik}, \citenamefont {Sen}, \citenamefont {Jha},\ and\
  \citenamefont {Mishra}}]{Patra2020}%
  \BibitemOpen
  \bibfield  {author} {\bibinfo {author} {\bibfnamefont {N.~K.}\ \bibnamefont
  {Patra}}, \bibinfo {author} {\bibfnamefont {T.}~\bibnamefont {Malik}},
  \bibinfo {author} {\bibfnamefont {D.}~\bibnamefont {Sen}}, \bibinfo {author}
  {\bibfnamefont {T.~K.}\ \bibnamefont {Jha}}, \ and\ \bibinfo {author}
  {\bibfnamefont {H.}~\bibnamefont {Mishra}},\ }\href {\doibase
  10.3847/1538-4357/aba8fc} {\bibfield  {journal} {\bibinfo  {journal}
  {Astrophys. J.}\ }\textbf {\bibinfo {volume} {900}},\ \bibinfo {pages} {49}
  (\bibinfo {year} {2020})}\BibitemShut {NoStop}%
\bibitem [{\citenamefont {Li}\ \emph {et~al.}(1998)\citenamefont {Li},
  \citenamefont {Ko},\ and\ \citenamefont {Bauer}}]{Li:1997px}%
  \BibitemOpen
  \bibfield  {author} {\bibinfo {author} {\bibfnamefont {B.-A.}\ \bibnamefont
  {Li}}, \bibinfo {author} {\bibfnamefont {C.~M.}\ \bibnamefont {Ko}}, \ and\
  \bibinfo {author} {\bibfnamefont {W.}~\bibnamefont {Bauer}},\ }\href
  {\doibase 10.1142/S0218301398000087} {\bibfield  {journal} {\bibinfo
  {journal} {Int. J. Mod. Phys. E}\ }\textbf {\bibinfo {volume} {7}},\ \bibinfo
  {pages} {147} (\bibinfo {year} {1998})},\ \Eprint
  {http://arxiv.org/abs/nucl-th/9707014} {arXiv:nucl-th/9707014} \BibitemShut
  {NoStop}%
\bibitem [{\citenamefont {Kurkela}(2022)}]{Kurkela:2022elj}%
  \BibitemOpen
  \bibfield  {author} {\bibinfo {author} {\bibfnamefont {A.}~\bibnamefont
  {Kurkela}},\ }in\ \href@noop {} {\emph {\bibinfo {booktitle} {{15th
  Conference on Quark Confinement and the Hadron Spectrum}}}}\ (\bibinfo {year}
  {2022})\ \Eprint {http://arxiv.org/abs/2211.11414} {arXiv:2211.11414
  [hep-ph]} \BibitemShut {NoStop}%
\bibitem [{\citenamefont {Gorda}\ \emph {et~al.}(2022)\citenamefont {Gorda},
  \citenamefont {Komoltsev},\ and\ \citenamefont {Kurkela}}]{Gorda:2022jvk}%
  \BibitemOpen
  \bibfield  {author} {\bibinfo {author} {\bibfnamefont {T.}~\bibnamefont
  {Gorda}}, \bibinfo {author} {\bibfnamefont {O.}~\bibnamefont {Komoltsev}}, \
  and\ \bibinfo {author} {\bibfnamefont {A.}~\bibnamefont {Kurkela}},\
  }\href@noop {} {\  (\bibinfo {year} {2022})},\ \Eprint
  {http://arxiv.org/abs/2204.11877} {arXiv:2204.11877 [nucl-th]} \BibitemShut
  {NoStop}%
\bibitem [{\citenamefont {Dong}\ \emph {et~al.}(2015)\citenamefont {Dong},
  \citenamefont {Zuo},\ and\ \citenamefont {Gu}}]{Dong:2015vga}%
  \BibitemOpen
  \bibfield  {author} {\bibinfo {author} {\bibfnamefont {J.}~\bibnamefont
  {Dong}}, \bibinfo {author} {\bibfnamefont {W.}~\bibnamefont {Zuo}}, \ and\
  \bibinfo {author} {\bibfnamefont {J.}~\bibnamefont {Gu}},\ }\href {\doibase
  10.1103/PhysRevC.91.034315} {\bibfield  {journal} {\bibinfo  {journal} {Phys.
  Rev. C}\ }\textbf {\bibinfo {volume} {91}},\ \bibinfo {pages} {034315}
  (\bibinfo {year} {2015})},\ \Eprint {http://arxiv.org/abs/1504.02217}
  {arXiv:1504.02217 [nucl-th]} \BibitemShut {NoStop}%
\bibitem [{\citenamefont {Chen}\ \emph {et~al.}(2009)\citenamefont {Chen},
  \citenamefont {Cai}, \citenamefont {Ko}, \citenamefont {Li}, \citenamefont
  {Shen},\ and\ \citenamefont {Xu}}]{Chen:2009wv}%
  \BibitemOpen
  \bibfield  {author} {\bibinfo {author} {\bibfnamefont {L.-W.}\ \bibnamefont
  {Chen}}, \bibinfo {author} {\bibfnamefont {B.-J.}\ \bibnamefont {Cai}},
  \bibinfo {author} {\bibfnamefont {C.~M.}\ \bibnamefont {Ko}}, \bibinfo
  {author} {\bibfnamefont {B.-A.}\ \bibnamefont {Li}}, \bibinfo {author}
  {\bibfnamefont {C.}~\bibnamefont {Shen}}, \ and\ \bibinfo {author}
  {\bibfnamefont {J.}~\bibnamefont {Xu}},\ }\href {\doibase
  10.1103/PhysRevC.80.014322} {\bibfield  {journal} {\bibinfo  {journal} {Phys.
  Rev. C}\ }\textbf {\bibinfo {volume} {80}},\ \bibinfo {pages} {014322}
  (\bibinfo {year} {2009})},\ \Eprint {http://arxiv.org/abs/0905.4323}
  {arXiv:0905.4323 [nucl-th]} \BibitemShut {NoStop}%
\bibitem [{\citenamefont {Fattoyev}\ \emph {et~al.}(2010)\citenamefont
  {Fattoyev}, \citenamefont {Horowitz}, \citenamefont {Piekarewicz},\ and\
  \citenamefont {Shen}}]{Fattoyev:2010mx}%
  \BibitemOpen
  \bibfield  {author} {\bibinfo {author} {\bibfnamefont {F.~J.}\ \bibnamefont
  {Fattoyev}}, \bibinfo {author} {\bibfnamefont {C.~J.}\ \bibnamefont
  {Horowitz}}, \bibinfo {author} {\bibfnamefont {J.}~\bibnamefont
  {Piekarewicz}}, \ and\ \bibinfo {author} {\bibfnamefont {G.}~\bibnamefont
  {Shen}},\ }\href {\doibase 10.1103/PhysRevC.82.055803} {\bibfield  {journal}
  {\bibinfo  {journal} {Phys. Rev. C}\ }\textbf {\bibinfo {volume} {82}},\
  \bibinfo {pages} {055803} (\bibinfo {year} {2010})},\ \Eprint
  {http://arxiv.org/abs/1008.3030} {arXiv:1008.3030 [nucl-th]} \BibitemShut
  {NoStop}%
\bibitem [{\citenamefont {Agrawal}\ \emph {et~al.}(2012)\citenamefont
  {Agrawal}, \citenamefont {Sulaksono},\ and\ \citenamefont
  {Reinhard}}]{Agrawal:2012rx}%
  \BibitemOpen
  \bibfield  {author} {\bibinfo {author} {\bibfnamefont {B.~K.}\ \bibnamefont
  {Agrawal}}, \bibinfo {author} {\bibfnamefont {A.}~\bibnamefont {Sulaksono}},
  \ and\ \bibinfo {author} {\bibfnamefont {P.~G.}\ \bibnamefont {Reinhard}},\
  }\href {\doibase 10.1016/j.nuclphysa.2012.03.004} {\bibfield  {journal}
  {\bibinfo  {journal} {Nucl. Phys. A}\ }\textbf {\bibinfo {volume} {882}},\
  \bibinfo {pages} {1} (\bibinfo {year} {2012})},\ \Eprint
  {http://arxiv.org/abs/1204.2644} {arXiv:1204.2644 [nucl-th]} \BibitemShut
  {NoStop}%
\bibitem [{\citenamefont {Imam}\ \emph {et~al.}(2022)\citenamefont {Imam},
  \citenamefont {Patra}, \citenamefont {Mondal}, \citenamefont {Malik},\ and\
  \citenamefont {Agrawal}}]{Imam:2021dbe}%
  \BibitemOpen
  \bibfield  {author} {\bibinfo {author} {\bibfnamefont {S.~M.~A.}\
  \bibnamefont {Imam}}, \bibinfo {author} {\bibfnamefont {N.~K.}\ \bibnamefont
  {Patra}}, \bibinfo {author} {\bibfnamefont {C.}~\bibnamefont {Mondal}},
  \bibinfo {author} {\bibfnamefont {T.}~\bibnamefont {Malik}}, \ and\ \bibinfo
  {author} {\bibfnamefont {B.~K.}\ \bibnamefont {Agrawal}},\ }\href {\doibase
  10.1103/PhysRevC.105.015806} {\bibfield  {journal} {\bibinfo  {journal}
  {Phys. Rev. C}\ }\textbf {\bibinfo {volume} {105}},\ \bibinfo {pages}
  {015806} (\bibinfo {year} {2022})},\ \Eprint
  {http://arxiv.org/abs/2110.15776} {arXiv:2110.15776 [nucl-th]} \BibitemShut
  {NoStop}%
\bibitem [{\citenamefont {Mondal}\ and\ \citenamefont
  {Gulminelli}(2022)}]{Mondal:2021vzt}%
  \BibitemOpen
  \bibfield  {author} {\bibinfo {author} {\bibfnamefont {C.}~\bibnamefont
  {Mondal}}\ and\ \bibinfo {author} {\bibfnamefont {F.}~\bibnamefont
  {Gulminelli}},\ }\href {\doibase 10.1103/PhysRevD.105.083016} {\bibfield
  {journal} {\bibinfo  {journal} {Phys. Rev. D}\ }\textbf {\bibinfo {volume}
  {105}},\ \bibinfo {pages} {083016} (\bibinfo {year} {2022})},\ \Eprint
  {http://arxiv.org/abs/2111.04520} {arXiv:2111.04520 [nucl-th]} \BibitemShut
  {NoStop}%
\bibitem [{\citenamefont {Gezerlis}\ and\ \citenamefont
  {Carlson}(2010)}]{Gezerlis:2009iw}%
  \BibitemOpen
  \bibfield  {author} {\bibinfo {author} {\bibfnamefont {A.}~\bibnamefont
  {Gezerlis}}\ and\ \bibinfo {author} {\bibfnamefont {J.}~\bibnamefont
  {Carlson}},\ }\href {\doibase 10.1103/PhysRevC.81.025803} {\bibfield
  {journal} {\bibinfo  {journal} {Phys. Rev. C}\ }\textbf {\bibinfo {volume}
  {81}},\ \bibinfo {pages} {025803} (\bibinfo {year} {2010})},\ \Eprint
  {http://arxiv.org/abs/0911.3907} {arXiv:0911.3907 [nucl-th]} \BibitemShut
  {NoStop}%
\bibitem [{\citenamefont {Hebeler}\ \emph {et~al.}(2013)\citenamefont
  {Hebeler}, \citenamefont {Lattimer}, \citenamefont {Pethick},\ and\
  \citenamefont {Schwenk}}]{Hebeler:2013nza}%
  \BibitemOpen
  \bibfield  {author} {\bibinfo {author} {\bibfnamefont {K.}~\bibnamefont
  {Hebeler}}, \bibinfo {author} {\bibfnamefont {J.~M.}\ \bibnamefont
  {Lattimer}}, \bibinfo {author} {\bibfnamefont {C.~J.}\ \bibnamefont
  {Pethick}}, \ and\ \bibinfo {author} {\bibfnamefont {A.}~\bibnamefont
  {Schwenk}},\ }\href {\doibase 10.1088/0004-637X/773/1/11} {\bibfield
  {journal} {\bibinfo  {journal} {Astrophys. J.}\ }\textbf {\bibinfo {volume}
  {773}},\ \bibinfo {pages} {11} (\bibinfo {year} {2013})},\ \Eprint
  {http://arxiv.org/abs/1303.4662} {arXiv:1303.4662 [astro-ph.SR]} \BibitemShut
  {NoStop}%
\bibitem [{\citenamefont {Danielewicz}\ and\ \citenamefont
  {Lee}(2014)}]{Danielewicz:2013upa}%
  \BibitemOpen
  \bibfield  {author} {\bibinfo {author} {\bibfnamefont {P.}~\bibnamefont
  {Danielewicz}}\ and\ \bibinfo {author} {\bibfnamefont {J.}~\bibnamefont
  {Lee}},\ }\href {\doibase 10.1016/j.nuclphysa.2013.11.005} {\bibfield
  {journal} {\bibinfo  {journal} {Nucl. Phys. A}\ }\textbf {\bibinfo {volume}
  {922}},\ \bibinfo {pages} {1} (\bibinfo {year} {2014})},\ \Eprint
  {http://arxiv.org/abs/1307.4130} {arXiv:1307.4130 [nucl-th]} \BibitemShut
  {NoStop}%
\bibitem [{\citenamefont {Tsang}\ \emph {et~al.}(2009)\citenamefont {Tsang},
  \citenamefont {Zhang}, \citenamefont {Danielewicz}, \citenamefont {Famiano},
  \citenamefont {Li}, \citenamefont {Lynch},\ and\ \citenamefont
  {Steiner}}]{Tsang:2008fd}%
  \BibitemOpen
  \bibfield  {author} {\bibinfo {author} {\bibfnamefont {M.~B.}\ \bibnamefont
  {Tsang}}, \bibinfo {author} {\bibfnamefont {Y.}~\bibnamefont {Zhang}},
  \bibinfo {author} {\bibfnamefont {P.}~\bibnamefont {Danielewicz}}, \bibinfo
  {author} {\bibfnamefont {M.}~\bibnamefont {Famiano}}, \bibinfo {author}
  {\bibfnamefont {Z.}~\bibnamefont {Li}}, \bibinfo {author} {\bibfnamefont
  {W.~G.}\ \bibnamefont {Lynch}}, \ and\ \bibinfo {author} {\bibfnamefont
  {A.~W.}\ \bibnamefont {Steiner}},\ }\href {\doibase
  10.1103/PhysRevLett.102.122701} {\bibfield  {journal} {\bibinfo  {journal}
  {Phys. Rev. Lett.}\ }\textbf {\bibinfo {volume} {102}},\ \bibinfo {pages}
  {122701} (\bibinfo {year} {2009})},\ \Eprint {http://arxiv.org/abs/0811.3107}
  {arXiv:0811.3107 [nucl-ex]} \BibitemShut {NoStop}%
\bibitem [{\citenamefont {Danielewicz}\ \emph {et~al.}(2017)\citenamefont
  {Danielewicz}, \citenamefont {Singh},\ and\ \citenamefont
  {Lee}}]{Danielewicz:2016bgb}%
  \BibitemOpen
  \bibfield  {author} {\bibinfo {author} {\bibfnamefont {P.}~\bibnamefont
  {Danielewicz}}, \bibinfo {author} {\bibfnamefont {P.}~\bibnamefont {Singh}},
  \ and\ \bibinfo {author} {\bibfnamefont {J.}~\bibnamefont {Lee}},\ }\href
  {\doibase 10.1016/j.nuclphysa.2016.11.008} {\bibfield  {journal} {\bibinfo
  {journal} {Nucl. Phys. A}\ }\textbf {\bibinfo {volume} {958}},\ \bibinfo
  {pages} {147} (\bibinfo {year} {2017})},\ \Eprint
  {http://arxiv.org/abs/1611.01871} {arXiv:1611.01871 [nucl-th]} \BibitemShut
  {NoStop}%
\bibitem [{\citenamefont {Steiner}\ \emph {et~al.}(2010)\citenamefont
  {Steiner}, \citenamefont {Lattimer},\ and\ \citenamefont
  {Brown}}]{Steiner:2010fz}%
  \BibitemOpen
  \bibfield  {author} {\bibinfo {author} {\bibfnamefont {A.~W.}\ \bibnamefont
  {Steiner}}, \bibinfo {author} {\bibfnamefont {J.~M.}\ \bibnamefont
  {Lattimer}}, \ and\ \bibinfo {author} {\bibfnamefont {E.~F.}\ \bibnamefont
  {Brown}},\ }\href {\doibase 10.1088/0004-637X/722/1/33} {\bibfield  {journal}
  {\bibinfo  {journal} {Astrophys. J.}\ }\textbf {\bibinfo {volume} {722}},\
  \bibinfo {pages} {33} (\bibinfo {year} {2010})},\ \Eprint
  {http://arxiv.org/abs/1005.0811} {arXiv:1005.0811 [astro-ph.HE]} \BibitemShut
  {NoStop}%
\bibitem [{\citenamefont {Silich}\ and\ \citenamefont
  {Tenorio-Tagle}(2013)}]{Silich:2013qx}%
  \BibitemOpen
  \bibfield  {author} {\bibinfo {author} {\bibfnamefont {S.}~\bibnamefont
  {Silich}}\ and\ \bibinfo {author} {\bibfnamefont {G.}~\bibnamefont
  {Tenorio-Tagle}},\ }\href {\doibase 10.1088/0004-637X/765/1/43} {\bibfield
  {journal} {\bibinfo  {journal} {Astrophys. J.}\ }\textbf {\bibinfo {volume}
  {765}},\ \bibinfo {pages} {43} (\bibinfo {year} {2013})},\ \Eprint
  {http://arxiv.org/abs/1301.2619} {arXiv:1301.2619 [astro-ph.GA]} \BibitemShut
  {NoStop}%
\bibitem [{\citenamefont {Chen}\ \emph {et~al.}(2010)\citenamefont {Chen},
  \citenamefont {Ko}, \citenamefont {Li},\ and\ \citenamefont
  {Xu}}]{Chen:2010qx}%
  \BibitemOpen
  \bibfield  {author} {\bibinfo {author} {\bibfnamefont {L.-W.}\ \bibnamefont
  {Chen}}, \bibinfo {author} {\bibfnamefont {C.~M.}\ \bibnamefont {Ko}},
  \bibinfo {author} {\bibfnamefont {B.-A.}\ \bibnamefont {Li}}, \ and\ \bibinfo
  {author} {\bibfnamefont {J.}~\bibnamefont {Xu}},\ }\href {\doibase
  10.1103/PhysRevC.82.024321} {\bibfield  {journal} {\bibinfo  {journal} {Phys.
  Rev. C}\ }\textbf {\bibinfo {volume} {82}},\ \bibinfo {pages} {024321}
  (\bibinfo {year} {2010})},\ \Eprint {http://arxiv.org/abs/1004.4672}
  {arXiv:1004.4672 [nucl-th]} \BibitemShut {NoStop}%
\bibitem [{\citenamefont {Malik}\ \emph {et~al.}(2023)\citenamefont {Malik},
  \citenamefont {Ferreira},\ and\ \citenamefont
  {Provid\^encia}}]{Malik:2023mnx}%
  \BibitemOpen
  \bibfield  {author} {\bibinfo {author} {\bibfnamefont {T.}~\bibnamefont
  {Malik}}, \bibinfo {author} {\bibfnamefont {M.}~\bibnamefont {Ferreira}}, \
  and\ \bibinfo {author} {\bibfnamefont {C.}~\bibnamefont {Provid\^encia}},\
  }\href@noop {} {\  (\bibinfo {year} {2023})},\ \Eprint
  {http://arxiv.org/abs/2301.08169} {arXiv:2301.08169 [nucl-th]} \BibitemShut
  {NoStop}%
\bibitem [{\citenamefont {Dutra}\ \emph {et~al.}(2014)\citenamefont {Dutra},
  \citenamefont {Louren{\c{c}}o}, \citenamefont {Avancini}, \citenamefont
  {Carlson}, \citenamefont {Delfino}, \citenamefont {Menezes}, \citenamefont
  {Provid{\^e}ncia}, \citenamefont {Typel},\ and\ \citenamefont
  {Stone}}]{dutra2014relativistic}%
  \BibitemOpen
  \bibfield  {author} {\bibinfo {author} {\bibfnamefont {M.}~\bibnamefont
  {Dutra}}, \bibinfo {author} {\bibfnamefont {O.}~\bibnamefont
  {Louren{\c{c}}o}}, \bibinfo {author} {\bibfnamefont {S.}~\bibnamefont
  {Avancini}}, \bibinfo {author} {\bibfnamefont {B.}~\bibnamefont {Carlson}},
  \bibinfo {author} {\bibfnamefont {A.}~\bibnamefont {Delfino}}, \bibinfo
  {author} {\bibfnamefont {D.}~\bibnamefont {Menezes}}, \bibinfo {author}
  {\bibfnamefont {C.}~\bibnamefont {Provid{\^e}ncia}}, \bibinfo {author}
  {\bibfnamefont {S.}~\bibnamefont {Typel}}, \ and\ \bibinfo {author}
  {\bibfnamefont {J.}~\bibnamefont {Stone}},\ }\href@noop {} {\bibfield
  {journal} {\bibinfo  {journal} {Phy. Rev. C}\ }\textbf {\bibinfo {volume}
  {90}},\ \bibinfo {pages} {055203} (\bibinfo {year} {2014})}\BibitemShut
  {NoStop}%
\bibitem [{\citenamefont {Patra}\ \emph {et~al.}(2022)\citenamefont {Patra},
  \citenamefont {Imam}, \citenamefont {Agrawal}, \citenamefont {Mukherjee},\
  and\ \citenamefont {Malik}}]{Patra:2022yqc}%
  \BibitemOpen
  \bibfield  {author} {\bibinfo {author} {\bibfnamefont {N.~K.}\ \bibnamefont
  {Patra}}, \bibinfo {author} {\bibfnamefont {S.~M.~A.}\ \bibnamefont {Imam}},
  \bibinfo {author} {\bibfnamefont {B.~K.}\ \bibnamefont {Agrawal}}, \bibinfo
  {author} {\bibfnamefont {A.}~\bibnamefont {Mukherjee}}, \ and\ \bibinfo
  {author} {\bibfnamefont {T.}~\bibnamefont {Malik}},\ }\href {\doibase
  10.1103/PhysRevD.106.043024} {\bibfield  {journal} {\bibinfo  {journal}
  {Phys. Rev. D}\ }\textbf {\bibinfo {volume} {106}},\ \bibinfo {pages}
  {043024} (\bibinfo {year} {2022})},\ \Eprint
  {http://arxiv.org/abs/2203.08521} {arXiv:2203.08521 [nucl-th]} \BibitemShut
  {NoStop}%
\bibitem [{\citenamefont {Blaizot}\ \emph {et~al.}(1995)\citenamefont
  {Blaizot}, \citenamefont {Berger}, \citenamefont {Dechargé},\ and\
  \citenamefont {Girod}}]{BLAIZOT1995}%
  \BibitemOpen
  \bibfield  {author} {\bibinfo {author} {\bibfnamefont {J.}~\bibnamefont
  {Blaizot}}, \bibinfo {author} {\bibfnamefont {J.}~\bibnamefont {Berger}},
  \bibinfo {author} {\bibfnamefont {J.}~\bibnamefont {Dechargé}}, \ and\
  \bibinfo {author} {\bibfnamefont {M.}~\bibnamefont {Girod}},\ }\href
  {\doibase https://doi.org/10.1016/0375-9474(95)00294-B} {\bibfield  {journal}
  {\bibinfo  {journal} {Nucl. Phys. A}\ }\textbf {\bibinfo {volume} {591}},\
  \bibinfo {pages} {435} (\bibinfo {year} {1995})}\BibitemShut {NoStop}%
\bibitem [{\citenamefont {Blaizot}(1980)}]{Blaizot:1980tw}%
  \BibitemOpen
  \bibfield  {author} {\bibinfo {author} {\bibfnamefont {J.~P.}\ \bibnamefont
  {Blaizot}},\ }\href {\doibase 10.1016/0370-1573(80)90001-0} {\bibfield
  {journal} {\bibinfo  {journal} {Phys. Rept.}\ }\textbf {\bibinfo {volume}
  {64}},\ \bibinfo {pages} {171} (\bibinfo {year} {1980})}\BibitemShut
  {NoStop}%
\bibitem [{\citenamefont {Youngblood}\ \emph {et~al.}(1999)\citenamefont
  {Youngblood}, \citenamefont {Clark},\ and\ \citenamefont
  {Lui}}]{Youngblood:1999zza}%
  \BibitemOpen
  \bibfield  {author} {\bibinfo {author} {\bibfnamefont {D.~H.}\ \bibnamefont
  {Youngblood}}, \bibinfo {author} {\bibfnamefont {H.~L.}\ \bibnamefont
  {Clark}}, \ and\ \bibinfo {author} {\bibfnamefont {Y.~W.}\ \bibnamefont
  {Lui}},\ }\href {\doibase 10.1103/PhysRevLett.82.691} {\bibfield  {journal}
  {\bibinfo  {journal} {Phys. Rev. Lett.}\ }\textbf {\bibinfo {volume} {82}},\
  \bibinfo {pages} {691} (\bibinfo {year} {1999})}\BibitemShut {NoStop}%
\bibitem [{\citenamefont {Shlomo}\ \emph {et~al.}(2006)\citenamefont {Shlomo},
  \citenamefont {Kolomietz},\ and\ \citenamefont {Colo}}]{shlomo2006}%
  \BibitemOpen
  \bibfield  {author} {\bibinfo {author} {\bibfnamefont {S.}~\bibnamefont
  {Shlomo}}, \bibinfo {author} {\bibfnamefont {V.}~\bibnamefont {Kolomietz}}, \
  and\ \bibinfo {author} {\bibfnamefont {G.}~\bibnamefont {Colo}},\ }\href
  {\doibase 10.1140/epja/i2006-10100-3} {\bibfield  {journal} {\bibinfo
  {journal} {Eur. Phys. J. A}\ }\textbf {\bibinfo {volume} {30}},\ \bibinfo
  {pages} {23} (\bibinfo {year} {2006})}\BibitemShut {NoStop}%
\bibitem [{\citenamefont {Stone}\ \emph {et~al.}(2014)\citenamefont {Stone},
  \citenamefont {Stone},\ and\ \citenamefont {Moszkowski}}]{Stone:2014wza}%
  \BibitemOpen
  \bibfield  {author} {\bibinfo {author} {\bibfnamefont {J.~R.}\ \bibnamefont
  {Stone}}, \bibinfo {author} {\bibfnamefont {N.~J.}\ \bibnamefont {Stone}}, \
  and\ \bibinfo {author} {\bibfnamefont {S.~A.}\ \bibnamefont {Moszkowski}},\
  }\href {\doibase 10.1103/PhysRevC.89.044316} {\bibfield  {journal} {\bibinfo
  {journal} {Phys. Rev. C}\ }\textbf {\bibinfo {volume} {89}},\ \bibinfo
  {pages} {044316} (\bibinfo {year} {2014})},\ \Eprint
  {http://arxiv.org/abs/1404.0744} {arXiv:1404.0744 [nucl-th]} \BibitemShut
  {NoStop}%
\bibitem [{\citenamefont {Pearson}\ \emph {et~al.}(2010)\citenamefont
  {Pearson}, \citenamefont {Chamel},\ and\ \citenamefont
  {Goriely}}]{Pearson:2010zz}%
  \BibitemOpen
  \bibfield  {author} {\bibinfo {author} {\bibfnamefont {J.~M.}\ \bibnamefont
  {Pearson}}, \bibinfo {author} {\bibfnamefont {N.}~\bibnamefont {Chamel}}, \
  and\ \bibinfo {author} {\bibfnamefont {S.}~\bibnamefont {Goriely}},\ }\href
  {\doibase 10.1103/PhysRevC.82.037301} {\bibfield  {journal} {\bibinfo
  {journal} {Phys. Rev. C}\ }\textbf {\bibinfo {volume} {82}},\ \bibinfo
  {pages} {037301} (\bibinfo {year} {2010})},\ \Eprint
  {http://arxiv.org/abs/1009.3816} {arXiv:1009.3816 [nucl-th]} \BibitemShut
  {NoStop}%
\bibitem [{\citenamefont {Sagawa}\ \emph {et~al.}(2007)\citenamefont {Sagawa},
  \citenamefont {Yoshida}, \citenamefont {Zeng}, \citenamefont {Gu},\ and\
  \citenamefont {Zhang}}]{Sagawa:2007sp}%
  \BibitemOpen
  \bibfield  {author} {\bibinfo {author} {\bibfnamefont {H.}~\bibnamefont
  {Sagawa}}, \bibinfo {author} {\bibfnamefont {S.}~\bibnamefont {Yoshida}},
  \bibinfo {author} {\bibfnamefont {G.-M.}\ \bibnamefont {Zeng}}, \bibinfo
  {author} {\bibfnamefont {J.-Z.}\ \bibnamefont {Gu}}, \ and\ \bibinfo {author}
  {\bibfnamefont {X.-Z.}\ \bibnamefont {Zhang}},\ }\href {\doibase
  10.1103/PhysRevC.77.049902} {\bibfield  {journal} {\bibinfo  {journal} {Phys.
  Rev. C}\ }\textbf {\bibinfo {volume} {76}},\ \bibinfo {pages} {034327}
  (\bibinfo {year} {2007})},\ \Eprint {http://arxiv.org/abs/0706.0966}
  {arXiv:0706.0966 [nucl-th]} \BibitemShut {NoStop}%
\bibitem [{\citenamefont {Colo}\ \emph {et~al.}(2014)\citenamefont {Colo},
  \citenamefont {Garg},\ and\ \citenamefont {Sagawa}}]{Colo:2013yta}%
  \BibitemOpen
  \bibfield  {author} {\bibinfo {author} {\bibfnamefont {G.}~\bibnamefont
  {Colo}}, \bibinfo {author} {\bibfnamefont {U.}~\bibnamefont {Garg}}, \ and\
  \bibinfo {author} {\bibfnamefont {H.}~\bibnamefont {Sagawa}},\ }\href
  {\doibase 10.1140/epja/i2014-14026-9} {\bibfield  {journal} {\bibinfo
  {journal} {Eur. Phys. J. A}\ }\textbf {\bibinfo {volume} {50}},\ \bibinfo
  {pages} {26} (\bibinfo {year} {2014})},\ \Eprint
  {http://arxiv.org/abs/1309.1572} {arXiv:1309.1572 [nucl-th]} \BibitemShut
  {NoStop}%
\bibitem [{\citenamefont {Chen}\ and\ \citenamefont
  {Piekarewicz}(2014)}]{chen2014building}%
  \BibitemOpen
  \bibfield  {author} {\bibinfo {author} {\bibfnamefont {W.-C.}\ \bibnamefont
  {Chen}}\ and\ \bibinfo {author} {\bibfnamefont {J.}~\bibnamefont
  {Piekarewicz}},\ }\href@noop {} {\bibfield  {journal} {\bibinfo  {journal}
  {Phys. Rev. C}\ }\textbf {\bibinfo {volume} {90}},\ \bibinfo {pages} {044305}
  (\bibinfo {year} {2014})}\BibitemShut {NoStop}%
\bibitem [{\citenamefont {Lalazissis}\ \emph {et~al.}(1997)\citenamefont
  {Lalazissis}, \citenamefont {K{\"o}nig},\ and\ \citenamefont
  {Ring}}]{lalazissis1997new}%
  \BibitemOpen
  \bibfield  {author} {\bibinfo {author} {\bibfnamefont {G.}~\bibnamefont
  {Lalazissis}}, \bibinfo {author} {\bibfnamefont {J.}~\bibnamefont
  {K{\"o}nig}}, \ and\ \bibinfo {author} {\bibfnamefont {P.}~\bibnamefont
  {Ring}},\ }\href@noop {} {\bibfield  {journal} {\bibinfo  {journal} {Phys.
  Rev. C}\ }\textbf {\bibinfo {volume} {55}},\ \bibinfo {pages} {540} (\bibinfo
  {year} {1997})}\BibitemShut {NoStop}%
\end{thebibliography}%

\end{document}